\documentclass[
reprint,
superscriptaddress,
amsmath,amssymb,
aps,
prb,
floatfix,
]{revtex4-2}

\usepackage{graphicx}%
\usepackage{dcolumn}%
\usepackage{bm}%
\usepackage{physics}
\usepackage{mathrsfs}
\usepackage[colorlinks,linkcolor=blue,anchorcolor=blue,citecolor=blue,breaklinks=true]{hyperref}
\usepackage[ruled,vlined,]{algorithm2e} 
\usepackage[normalem]{ulem} %
\usepackage{orcidlink}

\DontPrintSemicolon

\usepackage{amsfonts,amssymb}
\usepackage{textcomp}
\usepackage{float}
\usepackage{color}
\usepackage{tikz}
\usepackage{amsthm}
\usepackage{mathtools}
\usepackage{hyperref}
\hypersetup{
    colorlinks = true,
    urlcolor = blue
}
\usepackage{dsfont}
\usepackage{comment}
\newtheorem{theorem}{Theorem}[section]
\theoremstyle{definition}
\newtheorem{definition}{Definition}[section]
\newtheorem{lemma}[theorem]{Lemma}
\newtheorem{corollary}{Corollary}[theorem]

\newcommand{\sket}[1]{| #1 \rangle}
\newcommand{\anglelr}[1]{\left\langle #1 \right\rangle}
\newcommand{\sanglelr}[1]{\langle #1 \rangle}
\newcommand{\sbraket}[2]{\sanglelr{ #1 | #2 }}

\newcommand{\lr}[1]{\left(  #1  \right)}
\newcommand{\lrm}[1]{\left[  #1  \right]}
\newcommand{\lrg}[1]{\left\{  #1  \right\}}
\newcommand{\de}{\partial}

\newcommand{\mrm}[1]{\mathrm{#1}}
\newcommand{\bo}[1]{\mathbf{#1}}
\newcommand{\bog}[1]{\boldsymbol{#1}}

\newcommand{\dimer}[2]{\sigma^{\alpha_{#1 #2}}_{#1}\sigma^{\alpha_{#1 #2}}_{#2}}

\newcommand{\wordfig}[2]{\begin{minipage}[h]{#1 cm}
	\vspace{0pt}
	\includegraphics[width=\textwidth]{#2}
   \end{minipage}}

\newcommand{\lsegment}[2]{\mathcal{S}_{\overleftarrow{#1#2}}}
    
\newcommand{\TReven}{\bo{\mathsf{Ev}}}
\newcommand{\TRodd}{\bo{\mathsf{Od}}}
\newcommand{\Good}{\bo{\mathsf{Gd}}}

\begin{document}

\preprint{APS/123-QED}

\title{Exact Organization of Density Matrices and Entanglement Structure in the Kitaev Spin Liquid}

\author{Chen-Chih Wang, 
\orcidlink{0009-0006-6656-8071}}
\affiliation{Physics Department, National Tsing Hua University, Hsinchu 30013, Taiwan}

\author{Yi-Ping Huang, \orcidlink{0000-0001-6453-1191}}
\email[Correspondence email address: ]{yphuang@phys.nthu.edu.tw} 

\affiliation{Physics Department, National Tsing Hua University, Hsinchu 30013, Taiwan}
\affiliation{Physics Division, National Center for Theoretical Sciences, Taipei 10617, Taiwan}
\affiliation{Institute of Physics, Academia Sinica, Taipei 115, Taiwan}

\author{Sungkit Yip, \orcidlink{0000-0002-3542-5186}}
\email[Correspondence email address: ]{yip@phys.sinica.edu.tw} 

\affiliation{Institute of Physics, Academia Sinica, Taipei 115, Taiwan}
\affiliation{Institute of Atomic and Molecular Sciences, Academia Sinica, Taipei 10617, Taiwan}

\date{\today}

\begin{abstract}
We give an exact form of the density matrix for the spin-1/2 Kitaev spin liquid, expressed in terms of spin operators, and study the entanglement structure of the Kitaev honeycomb model within the spin framework. 
We show that the density matrix is naturally organized by equivalence classes of string operators associated with the underlying gauge structure of the model.
With the explicit form of the density matrix, plus the exact Gauss law of the emergent gauge theory and the exact 1-form Wilson symmetry in the Kitaev model, we demonstrate the existence of the underlying symmetry-resolved block-diagonal structure of the reduced density matrix, which gives rise to the extensive degeneracy in the entanglement spectrum. 
The block-diagonal structure is then proven to be responsible for the separability of the entanglement entropy into the gauge and matter parts.
Furthermore, we extend the formalism to subsystems with an odd number of lattice sites, revealing a relation between the entanglement spectrum and the fermion parity that is seldom mentioned in the literature.

\end{abstract}

\maketitle

\tableofcontents

\section{Introduction}

Quantum spin liquids (QSLs) represent quantum phases beyond the conventional Ginzburg--Landau paradigm, characterized not by local order parameters but by long-range quantum entanglement, fractionalized excitations, and emergent gauge structures~\cite{Savary-Balent_2017,Zhou-Kanoda-Ng_2017}. Unlike conventional ordered phases, whose essential properties can often be captured through spontaneous symmetry breaking, the defining characteristics of a QSL are encoded in the structure of its many-body quantum state. Consequently, much of our theoretical understanding of QSLs has been developed through their zero-temperature ground-state wave functions and the genuinely quantum structures contained within them.

Quantum entanglement provides one of the most powerful frameworks for extracting such structures~\cite{ZengChenZhouWen2019,Wen_2017,Laflorencie_2016,Amico-Fazio-Osterloh-Vedral_2008,Eisert-Cramer-Plenio_2010,Lyu-Chandorkar-Kapoor-Takei-Sorensen-Witczak-Krempa_2026}. For a pure ground state, the reduced density matrix (RDM) of a subsystem provides a complete characterization of its bipartite entanglement. Its eigenvalues determine the entanglement entropy (EE) and entanglement spectrum (ES), which have become central tools for identifying topological order~\cite{Levin-Wen_2006,Kitaev-Preskill_2006}, bulk--boundary correspondence~\cite{Li-Haldane,Li-Haldane-1,Li-Haldane-2}, and other manifestations of long-range quantum entanglement. This ground-state and entanglement-based perspective has led to remarkable progress in understanding QSLs at zero temperature.

At the same time, the success of this perspective also exposes a limitation of our present theoretical understanding rather than a fundamental limitation of the underlying physics. The characteristic quantum structures of QSLs are presently understood most sharply for pure ground states, where entanglement admits particularly direct characterizations. Once the state becomes mixed, as at finite temperature or in an open quantum system, several inequivalent measures of quantum and mixed-state entanglement are possible, but no single quantity plays the same general role as the bipartite EE of a pure state. This raises a broader question: which characteristic structures of a QSL can be identified directly at the level of its physical density matrix, which remains well defined for pure, mixed, thermal, and open-system states? A necessary first step toward addressing this broader question is to understand how the density matrix is organized already in an exactly solvable QSL at zero temperature.

This question also has an important operational aspect. While the density matrix specifies a quantum state, its expression in the physical operator algebra connects that state to physically accessible observables and operations. For a quantum spin system, these are naturally formulated in terms of the physical spin degrees of freedom. An explicit density matrix expressed in spin operators therefore provides a natural interface with quantum-information and quantum-resource frameworks, in which quantum properties are characterized relative to physically meaningful operations~\cite{Nielsen_Chuang_2010_sec8,Chitambar-Gour,Leone-Oliviero-Hamma_2022}. Thus, understanding a QSL at the level of its physical spin density matrix is relevant not only for extending its characterization beyond pure ground states, but also for identifying quantum structures that are operationally accessible in the underlying spin system.

The Kitaev honeycomb model~\cite{Kitaev_2006} provides an ideal setting in which to establish such a physical density-matrix description. As one of the few exactly solvable interacting spin models realizing a $\mathbb{Z}_2$ QSL, it has become a paradigmatic platform for understanding fractionalization, emergent gauge fields, topological order, and anyonic excitations~\cite{Savary-Balent_2017,Zhou-Kanoda-Ng_2017}, with growing connections to candidate Kitaev materials and experiments~\cite{Matsuda-Shibauchi-Kee_2025,Kasahara-etal_2018,Sandilands-Tian-Plumb-Kim-Burch_2015,Nasu-Knolle-Kovrizhin-Motome-Moessner_2016}. Through the Majorana-fermion representation, the interacting spin problem is mapped to free Majorana fermions coupled to a static $\mathbb{Z}_2$ gauge field, providing a powerful framework for calculating its ground-state entanglement properties. In particular, the EE was shown to separate exactly into matter and gauge contributions~\cite{Yao-Qi_2010}, while the ES can be obtained from fermionic correlation functions using standard free-fermion techniques~\cite{Jin-Korepin_2004,Vidal-Latorre-Rico-Kitaev_2003,Latorre-Rico-Vidal_2004,Peschel_2003,Peschel-Eisler_2009,Chung-Peschel_2001,Cheong-Henley_2004}. These developments establish in considerable detail \emph{what} the entanglement properties of the Kitaev spin liquid are.

What remains less transparent is the physical operator organization underlying these results. The conventional entanglement construction is formulated in an enlarged Majorana Hilbert space with gauge redundancy, where the physical state is recovered after imposing gauge constraints and the gauge degrees of freedom intersecting the entanglement boundary require a special treatment~\cite{Yao-Qi_2010}. Consequently, although the EE and ES can be calculated exactly, their underlying structure is not manifest directly in the original spin Hilbert space. In particular, it remains desirable to understand, within a unified physical-spin formulation, how the density matrix is organized, what microscopic structure produces the matter--gauge separation of the EE, why the RDM develops its characteristic block structure and degeneracies, and how gauge constraints and boundary degrees of freedom are encoded directly in the physical density matrix. The recently emphasized role of $\mathbb{Z}_2$ one-form Wilson symmetry~\cite{Xu-Rakovszky-Knap-Pollmann_2025} further suggests that these structures may admit a direct symmetry-based interpretation. These questions concern not the computability of the known entanglement quantities, but the physical organization of the quantum state from which they arise.

In this work, we answer these questions by deriving the exact DM of the Kitaev quantum spin liquid entirely within the original spin Hilbert space. {We show that the DM, as well as the RDM, expressed in terms of spin-string operators naturally implies the separation between the matter and gauge sectors in the entanglement structure. Furthermore, we demonstrate that the RDM is organized into diagonal blocks arising from boundary degrees of freedom constrained jointly by Gauss's law and the exact $\mathbb{Z}_2$ 1-form Wilson symmetry recently discussed in Ref.~\cite{Xu-Rakovszky-Knap-Pollmann_2025}. This structure provides a microscopic origin of the characteristic degeneracy of the entanglement spectrum, identifying it directly with the gauge-constrained boundary degrees of freedom of the physical spin system. In particular, the method of evaluating EE proposed in Ref.~\cite{Yao-Qi_2010} was made possible by redefining the gauge degrees of freedom at the boundary, a technique however seeming applicable only when the subsystem contains an even number of sites since the Hilbert space of an odd number of Majorana fermions would be ill-defined. Our physical spin formulation resolves this ambiguity by treating the matter and gauge sectors on an equal footing. This provides a unified framework applicable to both even and odd subsystem sizes and clarifies the subtle role of subsystem fermion parity in EE that remains hidden in the standard Majorana representation.
}

Beyond providing an exact spin representation of the DM, our work establishes a microscopic organization of the physical quantum state that complements the established Majorana description. More broadly, it identifies the DM itself as a natural object for revealing structural properties of quantum spin liquids. We expect this perspective to provide a useful foundation for investigating which characteristic structures identified at zero temperature survive, evolve, or disappear in finite-temperature and more general mixed-state settings, thereby contributing toward a more unified understanding of quantum spin liquids beyond the pure-ground-state framework.

The remainder of this paper is organized as follows. In Sec.~\ref{sec:model-intro}, we briefly review the Kitaev model and establish the notation. In Sec.~\ref{sec:density-matrix}, we derive the exact DM in the spin representation and demonstrate its organization by equivalence classes of spin-string operators. In Sec.~\ref{sec:entanglement-structures}, we derive the RDM, recover the previously known EE, identify the origin of its block-diagonal structure and entanglement-spectrum degeneracy, and discuss these results from the perspective of higher-form symmetry~\cite{Xu-Rakovszky-Knap-Pollmann_2025}. Finally, Sec.~\ref{sec:summary} summarizes our results and discusses possible future directions.

\section{An introduction to the Kitaev Model}\label{sec:model-intro}

Here, we review the Kitaev model and some of its fundamental properties, including the emergent gauge-theory structure obtained by representing the model in terms of Majorana fermions and the spin correlation functions. The gauge theory perspective will be used to clarify the microscopic origin of the entanglement structure. At the same time, the constraints on the spin correlation functions allow us to formulate the DM in an analyzable form, so that the entanglement structure can be elucidated in the spin representation.

The Hamiltonian of the Kitaev model reads~\cite{Kitaev_2006}
\begin{equation}\label{eq:model-ham}
    H = -\sum_{\anglelr{ij}}J_{\alpha_{ij}}\dimer{i}{j} \quad ;\quad\wordfig{1.6}{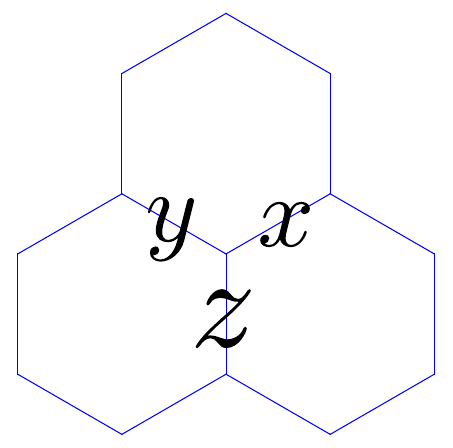}
\end{equation}
where $\sigma$ is the Pauli matrix representing a spin and $\alpha_{ij}=x,y,z$ depending on the direction of the edge $\anglelr{ij}$ shown in the figure above. It is convenient to use four types of Majorana fermions, also known as partons, $b^x$, $b^y$, $b^z$, and $c$, to represent the spin operators in the way that $\sigma^\mu = ib^\mu c$. Upon substitution, the spin Hamiltonian turns into a free fermion model coupled to a $\mathbb{Z}_2$ gauge field $\hat{u}_{ij}=ib^{\alpha_{ij}}_ib^{\alpha_{ij}}_j$ 
\begin{equation}\label{equ.MF-Hamiltonian}
    H \to \sum_{\anglelr{ij}}iJ_{\alpha_{ij}}\hat{u}_{ij}c_ic_j.
\end{equation}
The gauge field at each edge is a conserved variable that commutes with the Hamiltonian Eq.~\eqref{equ.MF-Hamiltonian}, and the ground state $\sket{\Tilde{\psi}} = \ket{u}\ket{M}$ can be expressed as a direct product of the gauge sector $\ket{u}$ and the matter sector $\ket{M}$, governed by the $b$-Majorana fermions and the $c$-Majorana fermions, respectively. However, the Majorana representation enlarges the original physical Hilbert space to a parton Hilbert space. To obtain the physical state $\ket{\psi}$, we have to apply a Gutzwiller projection $\mathcal{P}_G$ to project the state $\sket{\Tilde{\psi}}$ in the enlarged space to the physical state $\ket{\psi} = \mathcal{P}_G\sket{\Tilde{\psi}}$ in the physical space~\cite{Kitaev_2006, Pedrocchi-Chesi-Loss_2011, Zschocke-Vojta_2015}. Some basic information about the Gutzwiller projection in the Kitaev model is also introduced in Appendix~\ref{appendix:Majorana-derivation}.

Since the gauge field has no dynamics, the Hilbert space is divided into sectors labeled by the conserved fluxes, $W_p$, defined by
\begin{equation}\label{equ.flux-def}
    W_{p}\equiv \prod_{\anglelr{ij}\in p}\dimer{i}{j} \to \prod_{\anglelr{ij}\in p}\hat{u}_{ij},
\end{equation}
where $p$ represents an elementary hexagon of the honeycomb lattice.
The optimal configuration $\ket{u}$ of the gauge field that minimizes the ground state energy is the flux-free sector, in which $\anglelr{W_p}=+1\;\forall p$~\cite{Lieb_1994, Lieb-Loss_1992, Lieb_1992}. For convenience, we fix the gauge by choosing $u_{ij}=1\forall\anglelr{ij}$ such that the flux-free condition is satisfied. In this work, we assume this gauge by default, unless otherwise specified.

The previous part demonstrates that the Kitaev model can be viewed as a free fermion system coupled to a static $\mathbb{Z}_2$ gauge field.
The gauge connection in the gauge field that emerged from the fermionization of the Kitaev model is defined in a way that $\mrm{e}^{iA_{ij}} = \hat{u}_{ij}$, and hence the plaquette operator $W_p = \exp\lr{i\oint A}$ is the magnetic flux of the gauge theory. In contrast to the gauge field $A$, the electric field operators are defined in terms of the $b$-Majorana fermions in the way that $b^{\alpha_{ij}}_i = \mrm{e}^{i\pi E_{ij}}$ since they \textit{flip} the gauge fields according to the relation $b^{\alpha_{ij}}_i\hat{u}_{ij}b^{\alpha_{ij}}_i = -\hat{u}_{ij}$, which is analogous to the canonical quantization of gauge theories $[A_{ij}, E_{i'j'}] = -i\delta_{ii'}\delta_{jj'}$~\footnote{The quantization relation of the gauge theory depends on the gauge choice. In our discussion, the gauge is chosen as $A_0=0$, which leads to the relation $[A_\mu(\bo{x}), E_\nu(\bo{x}')] = -i\delta_{\mu\nu}\delta(\bo{x}-\bo{x}')$ mentioned in the main text.}. The Gauss law of the gauge theory only exists in the Majorana representation and reads
$b^x_ib^y_ib^z_ic_i = \mathds{1}$~\footnote{In this Gauss law formula, the $c$-Majorana fermion operator is exactly the matter field $c_i=\mrm{e}^{i\pi\rho_i}$ if the $b$-Majorana operators are represented as $b^{\alpha_{ij}}_i=\mrm{e}^{i\pi E_{ij}}$. After that, the exponent becomes the familiar form $\div{E}=\rho$}, which corresponds to the intrinsic algebraic relation $\sigma^x\sigma^y\sigma^z=i\mathds{1}$ of the Pauli matrices in the spin representation~\cite{Kitaev_2006}. {This constraint is also known as the Gutzwiller projection, since it projects states back to the physical sector as mentioned earlier. See Appendix~\ref{appendix:Majorana-derivation} and the references therein for details.}

The two-spin correlation in the Kitaev model is short-ranged. Only the correlation functions of two nearest-neighbor spins of the same type as the bond connecting them are finite, because two non-adjacent spin operators flip the gauge fluxes, which is orthogonal to the flux-free state~\cite{Bakaran-Mandal-Shankar_2007}. From this result, in general, only spin correlation functions of the form
\begin{equation}\label{equ.only-finite-correlation}
    \anglelr{\prod_{\anglelr{ij}\in \mathfrak{D}}\dimer{i}{j}}
\end{equation}
that consists of nearest-neighbor bonds survives. In the definition, we denote by $\mathfrak{D}$ a set of edges in the lattice.  Otherwise, the spin correlation function vanishes. Such a form of multi-spin correlation functions can be calculated by means of representing spin operators in terms of Majorana operators
\begin{equation}\label{equ.coefficient-Majorana-evaluation}
    \anglelr{\prod_{\anglelr{ij}\in \mathfrak{D}}\dimer{i}{j}} = \bra{M}\lr{\prod_{\anglelr{ij}\in\mathfrak{D}}ic_jc_i}\ket{M}.
\end{equation} 

Notice that Eq.~\eqref{equ.coefficient-Majorana-evaluation} becomes unity if the edges in $\mathfrak{D}$ form closed loops, since each Majorana fermion cancels that from its former edge one by one. Therefore, correlation functions of two edge configurations $\mathfrak{D}$ and $\mathfrak{D}'$ that differ from each other by a closed loop are equal. Hence, the two configurations are said to be \textit{equivalent}. This equivalence relation will give rise to an \textit{equivalence class structure}, which is crucial to formulating the DM.

\section{Density Matrices}\label{sec:density-matrix}

In this section, we show that the pattern of the DM that implies the separability of the entanglement proposed in~\cite{Yao-Qi_2010} is encoded in the \textit{equivalence classes} of the \textit{edge configurations} that describe \textit{strings}, both of which are introduced in Sec.~\ref{sec:density-matrix-whole}. With the basic notion of the strings, we directly derive an exact form of the DM in Sec.~\ref{sec:density-matrix-calculate}. Using a similar idea, the RDM is also derived in Sec.~\ref{sec:density-matrix-reduced}, which will be used as a foundation for the analysis of the entanglement structure. In Sec.~\ref{sec:Emergent-Fermions-at-the-Endpoints}, we show that the pattern of the DM also directly reflects the identity between the Majorana-fermion representation and the spin representation. This relation provides a foundation for the microscopic origin of the entanglement structure, especially for the separability of the entanglement into the gauge and matter parts.

\subsection{String Operators and Equivalence Classes}\label{sec:density-matrix-whole}

The information of a state is stored in the correlation functions. In the Kitaev model, the form~\eqref{equ.only-finite-correlation} implies that the finite correlation functions are determined by string-like objects $\mathfrak{D}$. Thus, we need to discuss how to manipulate the fundamental string objects to further analyze the DM. In this section, we introduce the notions of the \textit{string operators} and their \textit{equivalence classes} to describe the spin correlation functions in the Kitaev model. Both will be the building blocks for the DM. We begin by introducing the algebraic rule of operating edge configurations, and then point out the structure of the equivalence classes from the algebra.

\textit{Edge configurations}, denoted by $\mathfrak{D}$ as appeared in Eq.~\eqref{equ.only-finite-correlation}, are sets consisting of edges in the lattice. Their \textit{symmetric difference}, denoted by $\ominus$, is defined as
\[
\mathfrak{D}\ominus\mathfrak{D}' \equiv \mathfrak{D}\cup\mathfrak{D}'-\mathfrak{D}\cap\mathfrak{D}'
\]
Properties such as commutativity $\mathfrak{D}\ominus\mathfrak{D}' = \mathfrak{D}'\ominus\mathfrak{D}$ and associativity $\lr{\mathfrak{D}_1\ominus\mathfrak{D}_2}\ominus\mathfrak{D}_3 = \mathfrak{D}_1\ominus\lr{\mathfrak{D}_2\ominus\mathfrak{D}_3}$ can be verified directly. This algebra has three other properties exhibiting its $\mathbb{Z}_2$ nature: (i) $\varnothing\ominus\varnothing = \varnothing$, (ii) $\varnothing\ominus\mathfrak{D}=\mathfrak{D}$, and (iii) $\mathfrak{D}\ominus\mathfrak{D} = \varnothing$. These features enable it to describe the strings, which will be introduced later, since they are also $\mathbb{Z}_2$ objects that square to one. An example that illustrates the rule is given below
\[
\lr{\wordfig{1.6}{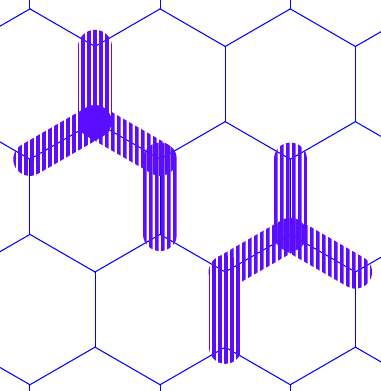}}
\ominus
\lr{\wordfig{1.6}{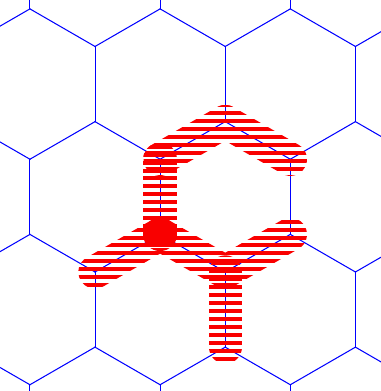}} 
= 
\lr{\wordfig{1.6}{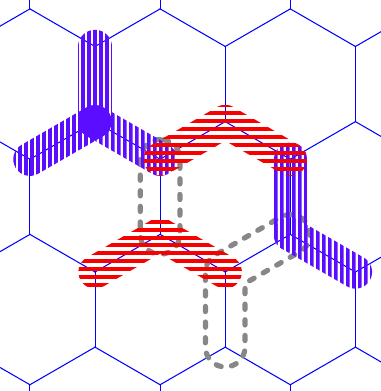}},
\]
where thick edges are those in the edge configurations, and the hollow edges with dotted boundaries on the right-hand side of the equation represent the annihilated edges since they belong to the intersection of the two string configurations. The algebra $\anglelr{\lrg{\mathfrak{D}},\ominus}$ is the fundamental constituent of the algebraic operations throughout this work. 

With the edge configurations, we define \emph{string operators} $\Sigma_\mathfrak{D}$ corresponding to different configurations $\mathfrak{D}$
\begin{equation}\label{equ.string-operator-def}
    \Sigma_\mathfrak{D} \equiv \prod_{\anglelr{ij}\in \mathfrak{D}}\sigma^{\alpha_{ij}}_i\sigma_j^{\alpha_{ij}}.
\end{equation}
The order of multiplying dimer operators $\dimer{i}{j}$ in the above product is also important, because two dimer operators anti-commute if they share a common endpoint. In a finite system, we can label the edges and multiply the dimer operators in order of this labeling to fix the ambiguity of the sign of the string operators. Due to the anti-commutation relation of Pauli spin operators, the string operators give a projective representation of the algebra $\anglelr{\lrg{\mathfrak{D}},\ominus}$ with a phase factor $\varphi(\mathfrak{D},\mathfrak{D}')=\pm 1$,
\begin{equation}\label{equ.Sigma-representation}
\Sigma_{\mathfrak{D}}\Sigma_{\mathfrak{D}'} = \varphi(\mathfrak{D},\mathfrak{D}') \Sigma_{\mathfrak{D}\ominus\mathfrak{D}'}.
\end{equation}
Detailed discussions on the fundamental properties of the phase factor $\varphi(\mathfrak{D},\mathfrak{D}')$ are given in Appendix~\ref{appendix:projective-rep}.

\begin{figure}
    \includegraphics[height=0.12\textheight]{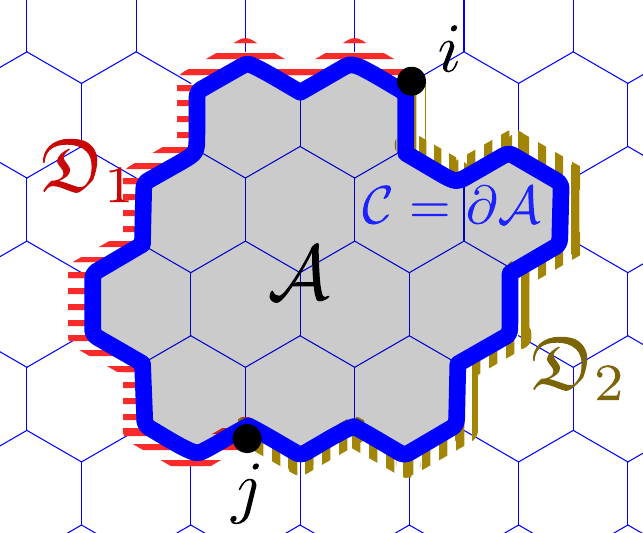}
    \hspace{3mm}
    \includegraphics[height=0.12\textheight]{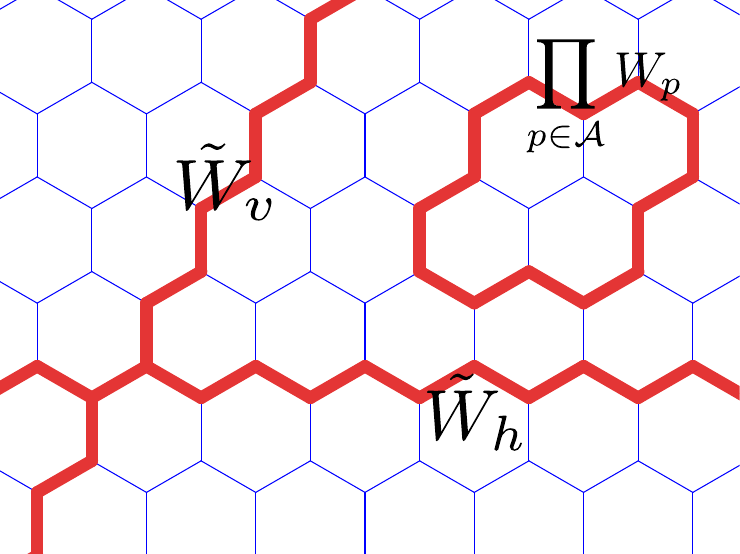}
    \caption{The left panel shows two equivalent edge configurations $\mathfrak{D}_1\sim\mathfrak{D}_2$ bridged via a loop $\mathcal{C}=\de\mathcal{A}$. The two configurations share the same endpoints $i$ and $j$. The right panel shows three possible ways to get other dimer configurations with the same expectation value of the string operators. The two Wilson loops, $\Tilde{W}_h\equiv \Sigma_{\mathcal{L}_h}$ and $\Tilde{W}_v\equiv \Sigma_{\mathcal{L}_v}$ commute $[\Tilde{W}_h,\Tilde{W}_v]=0$, and neither of them can be constructed by local operators $W_p$'s. Aside from the non-contractible Wilson loops, we also have local ones, given by plaquette operators $ W_p$ inside the region $\mathcal{A}$.}
    \label{fig.equivalent}
\end{figure}

Compared to the open strings with endpoints, edge configurations composed of closed loops form a special class, since the expectation values of their corresponding string operators are equal to 1. For example, the simplest closed loop $\de p$ consists of six edges surrounding a plaquette $p$, and the corresponding string operator 
\[
\Sigma_{\de p} = \prod_{\anglelr{ij}\in\de p}\sigma^{\alpha_{ij}}_i\sigma^{\alpha_{ij}}_j = W_p
\]
is a plaquette operator $W_p$ centered at $p$. From the property of $W_p$, string operators of larger loops, as well as edge configurations consisting of multiple loops, can be expressed as combinations of plaquette operators in a specific set $\mathcal{A}$ of plaquettes, in which the loop is the boundary of such a region
\begin{equation}\label{equ.lattice-Stokes}
    \Sigma_{\de\mathcal{A}} = \prod_{p\in\mathcal{A}}W_p,
\end{equation}
as illustrated in \autoref{fig.equivalent}. 
Since the Kitaev ground state is an eigenvector of plaquette operators with an eigenvalue $+1$, the expectation value of the string operator of the configuration $\mathfrak{D}$ is equal to another configuration $\mathfrak{D}\ominus\de\mathcal{A}$ constructed by adding a loop configuration $\de\mathcal{A}$ to the original one up to a sign $\varphi$
\[
\anglelr{\Sigma_{\de\mathcal{A}\ominus\mathfrak{D}}}
= 
\varphi(\de\mathcal{A},\mathfrak{D}) \anglelr{\Sigma_{\de\mathcal{A}}\Sigma_{\mathfrak{D}}} = 
\varphi(\de\mathcal{A},\mathfrak{D}) \anglelr{\Sigma_{\mathfrak{D}}}.
\]
Such an equation is equivalent to 
\begin{equation}\label{equ.one-loop-equivalence}
\lr{\prod_{p\in\mathcal{A}}W_p}
\anglelr{\Sigma_{\de\mathcal{A}\ominus\mathfrak{D}}}\Sigma_{\de\mathcal{A}\ominus\mathfrak{D}} = \anglelr{\Sigma_{\mathfrak{D}}}\Sigma_{\mathfrak{D}}
\end{equation}
for an arbitrary set $\mathcal{A}$ of plaquettes due to the $\mathbb{Z}_2$ nature of the string operators. Similarly, for the non-contractible loops $\mathcal{L}_v$ and $\mathcal{L}_h$ that circle the torus in the vertical and horizontal directions, we have
\begin{equation}\label{equ.global-loop-equivalence}
\Tilde{W}_{\mathcal{L}_{v/h}}\anglelr{\Sigma_{\mathcal{L}_{v/h}\ominus\mathfrak{D}}}\Sigma_{\mathcal{L}_{v/h}\ominus\mathfrak{D}} = \anglelr{\Sigma_{\mathfrak{D}}}\Sigma_{\mathfrak{D}}, 
\end{equation}
where $\Tilde{W}_v\equiv\Sigma_{\mathcal{L}_v}$ and $\Tilde{W}_h\equiv\Sigma_{\mathcal{L}_h}$ are the corresponding string operators of non-contractible loops $\mathcal{L}_v$ and $\mathcal{L}_h$, respectively, as illustrated in the right panel of \autoref{fig.equivalent}.
In summary, Eqs. Eq.~\eqref{equ.one-loop-equivalence} and Eq.~\eqref{equ.global-loop-equivalence} implies the following relation
\begin{equation}\label{equ.string-operator-equivalent}
\hat{\Pi}\anglelr{\Sigma_{\de\mathcal{A}\ominus\mathfrak{D}}}\Sigma_{\de\mathcal{A}\ominus\mathfrak{D}} 
= \hat{\Pi}\anglelr{\Sigma_{\mathfrak{D}}}\Sigma_{\mathfrak{D}},
\end{equation}
where 
\begin{equation}\label{eq.p-proj-def}
    \hat{\Pi} \equiv 
\lr{\frac{\mathds{1} + \Tilde{W}_v}{2}} 
\lr{\frac{\mathds{1} + \Tilde{W}_h}{2}}
\prod_{p} \lr{\frac{\mathds{1} + W_p}{2}}
\end{equation}
is a projector onto the space with the flux-free constraint $\anglelr{W_p}=+1\forall p$ and one of the choices for non-contractible Wilson loops $\sanglelr{\Tilde{W}_v} = \sanglelr{\Tilde{W}_h} = +1$~\footnote{For a torus, there are four choices in total for non-contractible Wilson loops corresponding to $\sanglelr{\Tilde{W}_v}=\pm 1$ with $\sanglelr{\Tilde{W}_h}=\pm 1$. These four cases are equivalent for the analysis of the entanglement feature, but only in different topological sectors. In this work, we only consider one of the topological sectors.}. In other words, the projector $\hat{\Pi}$ absorb a closed-loop string operator
\begin{equation}\label{eq.W_absorb}
    \hat{\Pi}\Sigma_{\mathcal{C}} = \hat{\Pi}
\end{equation}
for an arbitrary closed loop $\mathcal{C}$. Thus, only the information of configurations containing no loops remains. Relations Eq.~\eqref{equ.string-operator-equivalent} to Eq.~\eqref{eq.W_absorb} give rise to an equivalence relation for edge configurations: Two edge configurations $\mathfrak{D}_1$ and $\mathfrak{D}_2$ are said to be \emph{equivalent}, denoted by $\mathfrak{D}_1\sim\mathfrak{D}_2$, if and only if there exists a set $\mathcal{A}$ of plaquettes such that one of the following four conditions holds:
\begin{enumerate}
    \item[(i)] $\mathfrak{D}_1=\mathfrak{D}_2\ominus\de\mathcal{A}$
    \item[(ii)] $\mathfrak{D}_1=\mathfrak{D}_2\ominus\de\mathcal{A}\ominus\mathcal{L}_v$
    \item[(iii)] $\mathfrak{D}_1=\mathfrak{D}_2\ominus\de\mathcal{A}\ominus\mathcal{L}_h$
    \item[(iv)] $\mathfrak{D}_1=\mathfrak{D}_2\ominus\de\mathcal{A}\ominus\mathcal{L}_v\ominus\mathcal{L}_h$
\end{enumerate} 
Equivalent configurations are collected into \emph{equivalence classes} $[\mathfrak{D}^c]=\lrg{\mathfrak{D}:\mathfrak{D}\sim\mathfrak{D}^c}$ characterized by each representative $\mathfrak{D}^c$. At the same time, {the left panel} of Fig.~\ref{fig.equivalent} shows that the two equivalent edge configurations must share the same set of endpoints, providing a simple way to identify them. In other words, there is a one-to-one correspondence between an equivalence class for edge configurations $[\mathfrak{D}]$ and the \textit{set of endpoints} of the representative configuration $\mathfrak{D}$, denoted by $\mrm{EP}\lr{\mathfrak{D}}$. Aside from the convenience that endpoint sets uniquely determine the {equivalence class}, the physical picture of the equivalence classes and endpoint sets also implies that a Pauli string in the spin representation is identical to the two Majorana fermions residing at the endpoints of the string. This can be seen in the commutation relation among the endpoints of strings~\cite{Kitaev_2006, Kawagoe-Levin_2020, fermion-and-string}. {In fact, the physical interpretation of the properties of string operators listed from Eq.~\eqref{equ.lattice-Stokes} to Eq.~\eqref{eq.W_absorb} is that the system can be separated into the gauge and matter parts, in which the gauge part is governed by the projector $\hat{\Pi}$ while the matter part is encoded in the endpoints of string operators. The only difference is that this fact is demonstrated entirely using spin operators throughout this work. More details will be presented in Sec.~\ref{sec:Emergent-Fermions-at-the-Endpoints}.} 

\subsection{Density Matrix of the Kitaev Ground State on a Torus}\label{sec:density-matrix-calculate}

In Ref. \cite{Yao-Qi_2010}, the authors demonstrate the separability of EE between the gauge and matter sectors through the replica approach. However, as the Kitaev model can be exactly transformed to a free-fermion theory without disorder, we anticipate a direct demonstration of such separability without the replica approach. Here, we use the string operators as the building blocks to derive the DM of the Kitaev ground state. We illustrate that the separability between the two sectors arises from the equivalence class, and the representative of each equivalence class further reflects the origin of the effective Majorana fermion theory.

We start from the fundamental properties of the DM. A DM must be Hermitian, and it gives the expectation value of observables by $\anglelr{\mathcal{O}} = \Tr\lr{\rho \mathcal{O}}$. {Following this narrative, in general, a DM can be expanded by a set of orthonormal Hermitian basis operators $\lrg{\mathcal{O}_n}$ as follows}
\begin{equation}\label{equ.rho-starting}
    \rho = \frac{1}{2^N}\sum_{n}\anglelr{\mathcal{O}_n}\mathcal{O}_n,
\end{equation}
{where $N$ is the total number of spins in the system.} {In constructing the above equation, the operators $\mathcal{O}_n$ are chosen as Pauli group elements composed of Pauli spin operators, since every observable in a spin-1/2 system can be expanded in terms of Pauli group elements~\cite{Nielsen-Chuang_2010}. } For the Kitaev model, the property \eqref{equ.coefficient-Majorana-evaluation} constrains the form of finite observables composed of spin operators. Therefore, the expansion \eqref{equ.rho-starting} includes only terms of the form $\mathcal{O}_n = \Sigma_\mathfrak{D}$. Due to this, the density matrix is proportional to the following expansion
\[
\sum_{n} \anglelr{\mathcal{O}_n}\mathcal{O}_n
\to
\sum_{\mathfrak{D}} \mathfrak{C}_{\mathfrak{D}}\Sigma_\mathfrak{D}
\]
{We denote by $\mathfrak{C}_{\mathfrak{D}} \equiv \eta_{\mathfrak{D}}\anglelr{\Sigma_\mathfrak{D}}$ the expectation value of the string operators with a sign factor $\eta_D$. One can easily compute the expectation value of a string operator using the equivalent Majorana fermion formula Eq.~\eqref{equ.coefficient-Majorana-evaluation}. The sign factor $\eta_\mathfrak{D}$, which represents the sign obtained by squaring the string operator $\Sigma_\mathfrak{D}^2 = \eta_\mathfrak{D}\mathds{1}$, is introduced in Def.~\ref{def:Teven-odd} because certain string operators square to $-1$ rather than $+1$, as illustrated by Lemma~\ref{lemma:sigma-eta-relation}. Nevertheless, the sign factor $\eta_\mathfrak{D}$ can be neglected if we restrict ourselves to the ground state of the pure Kitaev model described by the Hamiltonian~\eqref{eq:model-ham} since the edge configurations that possess a finite coefficient $\mathfrak{C}_\mathfrak{D}$ satisfy $\eta_\mathfrak{D} = +1$, as explained in Corollary~\ref{corollary:good-square}. See Appendix~\ref{appendix:projective-rep} for further details.} 
The terms in the expansion \eqref{equ.rho-starting} can be sorted into equivalence classes since every edge configuration in the same equivalence class gives the same weight to the DM.
\begin{equation}\label{equ.class-sum}
    \sum_{\mathfrak{D}}\mathfrak{C}_{\mathfrak{D}}\Sigma_\mathfrak{D} = \sum_{[\mathfrak{D}^c]}\sum_{\mathfrak{D}\in[\mathfrak{D}^c]}\mathfrak{C}_{\mathfrak{D}}\Sigma_\mathfrak{D}.
\end{equation}
According to Eq.~\eqref{equ.string-operator-equivalent}, all terms in an equivalence class can be related to each other via the projector $\hat{\Pi}$. Thus, the summation in \eqref{equ.class-sum} can be factorized into groups by multiplying $\hat{\Pi}$. Therefore, we anticipate that the DM would be of the following form
\begin{equation}\label{equ.rho-factorization}
    \rho = C \hat{\Pi} \sum_{[\mathfrak{D}]} \mathfrak{C}_{\mathfrak{D}} \Sigma_{\mathfrak{D}}.
\end{equation}
The constant $C$ is the normalization factor determined by imposing the constraint $\Tr\rho = 1$. Performing the trace, we have 
\begin{align}\label{equ.rho-direct-trace-1}
    \Tr\rho &= {C} \sum_{[\mathfrak{D}]} \mathfrak{C}_{\mathfrak{D}} \Tr\lr{\hat{\Pi}\Sigma_{\mathfrak{D}}} \nonumber\\
    &= {\frac{1}{2^2}}\frac{C}{2^{N_p}} \sum_{[\mathfrak{D}]} \sum_{\mathcal{A}}'\Tr\lrm{ \lr{2\prod_{p\in\mathcal{A}} W_p}\Sigma_{\mathfrak{D}}},
\end{align}
{where $N_p$ represents the total number of hexagons in the lattice~\footnote{For a honeycomb lattice on a torus, the relationship between the total number of lattice sites and the total number of hexagons is $N = 2N_p$}.} The coefficient $1/2^2$ in the {second} line originates from the denominator of projectors provided by the two non-contractible Wilson loops. The two non-contractible Wilson-loop operators drop out because neither the local plaquette operators nor the string operators can annihilate them to yield the identity. Consequently, their contributions vanish upon taking the trace. The summation of plaquette operators arises from the expansion of $\hat{\Pi}$
\begin{align}\label{equ.W-prod-expansion}
    \prod_{p}\lr{\mathds{1}+W_p} &= \lr{\mathds{1}+\prod_{p}W_p}\sum'_{\mathcal{A}}\prod_{p\in\mathcal{A}}W_p \nonumber\\
    &= 2\sum'_{\mathcal{A}}\prod_{p\in\mathcal{A}}W_p,
\end{align}
with $\mathcal{A}$ denoting different sets of plaqette configurations defined in Eq.~\eqref{equ.lattice-Stokes}. We point out that, due to the identity $\prod_{p}W_p=\mathds{1}$, the loop constructed by $\prod_{p\in \mathcal{A}}W_p$ is identical to that constructed by $\prod_{p\in \overline{\mathcal{A}}}W_p$ with $\overline{\mathcal{A}}$ representing the complement of $\mathcal{A}$. To remove redundancy, we use the primed sum $\sum'$ in Eq.~\eqref{equ.W-prod-expansion} that sums over only independent sets of plaquettes. The only two terms that contribute to the trace in Eq.~\eqref{equ.rho-direct-trace-1} are (i) $\mathcal{A}=\varnothing$ and $\mathfrak{D}\sim \varnothing$ and (ii) $\mathcal{A}=\mathfrak{P}$ and $\mathfrak{D}\sim\mathfrak{D}_Z$, where the definitions of $\mathfrak{P}$ and $\mathfrak{D}_Z$ is introduced in Appendix~\ref{sec:fermion_parity_from_string_operators} and \autoref{fig.S-P-dimer-gas}. Therefore, Eq.~\eqref{equ.rho-direct-trace-1} is equal to 
\[
\Tr\rho = \frac{C}{2^{N_p}}\Tr\mathds{1} = \frac{C}{2^{-N+N_p}}.
\]
To make the trace equal to $1$, we require $C = 2^{-N+N_p}$. 
In the end, we obtain the exact form of the DM
\begin{equation}\label{equ.density-matrix}
    \rho =  \frac{1}{2^{N-N_p}}\hat{\Pi}\sum_{[\mathfrak{D}]}\mathfrak{C}_{\mathfrak{D}}\Sigma_{\mathfrak{D}},
\end{equation}
{with $N$ denoting the number of lattice sites and $N_p$ the number of hexagons (plaquettes) in the lattice}. As indicated in Sec.~\ref{sec:density-matrix-whole}, there is a one-to-one correspondence between sets of endpoints and equivalence classes. Therefore, the sum in Eq.~\eqref{equ.density-matrix} can also be understood as summing over sets of endpoints. This implies that the string details are smeared out, and only the information about their endpoints remains important. The information about string details is grouped and stored in the projector $\hat{\Pi}$.

\subsection{Reduced Density Matrix}\label{sec:density-matrix-reduced}

Understanding the RDM is the first step in analyzing the entanglement structure, as key quantities such as ES and EE are defined directly by the RDM. Here, we employ the method introduced in the previous section to obtain the RDM intuitively without performing the partial trace. Nevertheless, we still provide a rigorous derivation of how to perform the partial trace directly in Appendix~\ref{appendix:reduced-density0matrix}. 

Like the full DM of the entire system, a RDM should be Hermitian and yields the expectation values of subsystem observables via $\anglelr{\mathcal{O}_A}=\Tr\lr{\rho\mathcal{O}_A}=\Tr_A\lr{\rho_A\mathcal{O}_A}$ for any observable $\mathcal{O}_A$ support within $A$. In the Kitaev model, observables possessing finite expectation values are string operators. Consequently, the RDM can be expanded in terms of string operators $\Sigma_\mathfrak{D}$, with $\mathfrak{D}$ entirely within $A$, weighted by their respective expectation values $\mathfrak{C}_\mathfrak{D}$,
\begin{equation}\label{equ.rho-A-raw-guess}
    \rho_A \propto\sum_{\mathfrak{D}\subset A}\mathfrak{C}_\mathfrak{D} \Sigma_\mathfrak{D}
\end{equation}
The equivalent configurations can be sorted in the same way as in the previous section. Owing to this, the summation can be rewritten as a summation of {equivalence classes} with the projector $\hat{\Pi}_A$ of dimension $2^{N^A-N^A_p}$, {where $N^A$ represents the number of lattice sites enclosed in the subregion $A$, and $N^A_p$ is the number of hexagons that lie entirely inside the subregion $A$},
\begin{equation}\label{equ.p-proj-A}
    \hat{\Pi}_A \equiv \prod_{p\in A}\lr{\frac{\mathds{1}+W_p}{2}}.
\end{equation}
{The subscript $p\in A$ stands for including only plaquettes that lie entirely inside $A$.} The final form of the RDM is then
\[
\rho_A = C\hat{\Pi}_A\sum_{[\mathfrak{D}]\subset A}\mathfrak{C}_{\mathfrak{D}}\Sigma_{\mathfrak{D}}.
\]
The condition $[\mathfrak{D}] \subset A$ restricts the summation to equivalence classes containing at least one edge configuration $\mathfrak{D}$ supported entirely within $A$. This choice ensures that $\Sigma_{\mathfrak{D}}$ is well-defined and acts purely on the local Hilbert space of subsystem $A$. Recalling the correspondence between endpoint sets and {equivalence classes}, this requirement is also equivalent to selecting equivalence classes $[\mathfrak{D}]$ such that $\mrm{EP}(\mathfrak{D})\subset A$. The constant $C$ is a normalization coefficient to be determined by imposing the normalization condition $\Tr_A\rho_A=1$,
\begin{align*}
    \Tr_A\rho_A &= C\sum_{[\mathfrak{D}]\subset A} \mathfrak{C}_\mathfrak{D} \Tr_A\lrm{\hat{\Pi}_A \Sigma_\mathfrak{D}} \\
    &= \frac{C}{2^{N^A_p}} \sum_{{\mathcal{A} \subset A}} \sum_{[\mathfrak{D}]\subset A} \mathfrak{C}_\mathfrak{D} \Tr_A\lrm{ \lr{\prod_{p\in\mathcal{A}}W_p} \Sigma_\mathfrak{D}} \\
    &= \frac{C}{2^{N^A_p}}\Tr_A\mathds{1}_A = C2^{N^A - N^A_p} = 1
\end{align*}
Therefore, the final form of the RDM reads
\begin{equation}\label{equ.red-den-mat-sim-con}
    \rho_A = \frac{1}{2^{N^A-N^A_p}} \hat{\Pi}_A\sum_{[\mathfrak{D}]\subset A}\mathfrak{C}_{\mathfrak{D}}\Sigma_{\mathfrak{D}},
\end{equation}
{with $N^A$ representing the number of lattice sites enclosed in the subregion $A$ and $N^A_p$ the number of hexagons that lie entirely inside the subregion $A$}. As anticipated, the RDM \eqref{equ.red-den-mat-sim-con} takes the same form as the full DM in \eqref{equ.density-matrix}.

\subsection{Emergent Fermions at the Endpoints of Strings -- Equivalence between Spin and Fermion Representation}\label{sec:Emergent-Fermions-at-the-Endpoints}

The form of Eq.~\eqref{equ.density-matrix} and Eq.~\eqref{equ.red-den-mat-sim-con} implies the separability of the matter sector from the gauge sector via a pure spin representation. This observation also reflects the similarity between the DM represented using spins and the one using Majorana fermions. In this section, we demonstrate that the spin representation and the Majorana-fermion representation are identical, and they are just two different representations of the same algebra. Such an identity makes the DM reproduce correct EE and further implies the block-diagonal structure of the DM.

Starting from the Majorana representation, the ground state DM can also be represented as
\[
  \rho = 2^{N-1}\mathcal{P}_G\prod_{\anglelr{ij}}\lr{\frac{\mathds{1}+\hat{u}_{ij}}{2}}\rho_F\mathcal{P}_G,
\]
where $\rho_F$ is the DM of the Kitaev ground state in the Majorana fermion representation. The Majorana DM $\rho_F$ can be constructed from a starting point similar to Eq.~\eqref{equ.rho-starting} that we have used in the beginning,
\begin{equation}\label{equ.rho-F}
    \rho^F = \frac{1}{2^{N/2}}\sum_{[\mathfrak{D}]}\mathfrak{C}_\mathfrak{D}\mathcal{M}_\mathfrak{D},
\end{equation}
where
\begin{equation}\label{equ.V-rep-def}
    \mathcal{M}_\mathfrak{D} \equiv \prod_{\anglelr{ij}\in\mathfrak{D}}ic_jc_i.
\end{equation}
Details of the derivations are presented in Appendix \ref{appendix:Majorana-derivation}. In each dimer $\anglelr{ij}$ of the above product, the site $i$ belongs to sublattice $A$ and $j$ to sublattice $B$. The denominator of the normalization coefficient is $2^{N/2}$ instead of $2^N$ because a normal fermion $f$ is defined by two $c$-Majorana fermions
. To manifest the similarity between the spin DM \eqref{equ.density-matrix} and the Majorana DM \eqref{equ.rho-F}, we define a set of operators $\mathcal{S}_{\mathfrak{D}}$ for each operator that corresponds to an edge configuration
\begin{equation}\label{equ.S-def}
    \mathcal{S}_\mathfrak{D} \equiv \hat{\Pi}\Sigma_\mathfrak{D}.
\end{equation}
Given Eq.~\eqref{equ.Sigma-representation} and the fact that $[{\hat{\Pi}},\Sigma_{\mathfrak{D}}]=0$, the set of operators $\lrg{\mathcal{S}_{\mathfrak{D}}}$ also forms a representation of the algebra $\anglelr{\lrg{\mathfrak{D}},\ominus}$, 
\begin{equation}\label{equ.U-rep-algebra}
    \mathcal{S}_{\mathfrak{D}}\mathcal{S}_{\mathfrak{D}'} = \varphi(\mathfrak{D},\mathfrak{D}')\mathcal{S}_{\mathfrak{D}\ominus\mathfrak{D}'}.
\end{equation}
With the string representation \eqref{equ.S-def} and \eqref{equ.U-rep-algebra}, the form of the DM in Eq.~\eqref{equ.density-matrix} becomes
\begin{equation}\label{equ.rho-U}
    \rho =  \frac{1}{2^{N-N_p}} \sum_{[\mathfrak{D}]}\mathfrak{C}_{\mathfrak{D}}\mathcal{S}_{\mathfrak{D}},
\end{equation}
which is of a form similar to Eq.~\eqref{equ.rho-F} if those $\mathcal{S}$-operators are replaced by $\mathcal{M}$-operators. This is reminiscent of different representations of the same algebra.
If we use the Majorana fermions to represent spin operators as $\sigma^\mu\to ib^\mu c$, then 
\begin{align*}
    \mathcal{S}_\mathfrak{D} &\to \hat{\Pi} \prod_{\anglelr{ij}\in \mathfrak{D}}\lr{\hat{u}_{ij}ic_jc_i} \\
    &= \hat{\Pi} \lrm{\prod_{\anglelr{ij}\in \mathfrak{D}}\hat{u}_{ij}} \mathcal{M}_{\mathfrak{D}}
\end{align*}
The first two products completely consist of $\hat{u}_{ij}$'s, which commute with each other as well as the $c$-Majorana fermion operators~\footnote{Note that the projector $\hat{\Pi}$ becomes
\[\hat{\Pi} \to \prod_{ p}\lr{\frac{\mathds{1} + \prod_{\anglelr{ij}\in p}\hat{u}_{ij}}{2}}\]
in the Majorana representation, where ``$\anglelr{ij}\in p$'' stands for \textit{those edges of the plaquette $p$}.}. Therefore, they will not contribute to the phase factor $\varphi\lr{\mathfrak{D},\mathfrak{D}'}$ at all. So, we know that the anti-commutation relations among $c$-Majorana fermions are encoded in the phase factor $\varphi\lr{\mathfrak{D},\mathfrak{D}'}$. In view of this, we conclude that the operators $\mathcal{M}_\mathfrak{D}$ defined by Eq.~\eqref{equ.V-rep-def} also form a projective representation of the edge configuration algebra and share the same phase factor $\varphi\lr{\mathfrak{D},\mathfrak{D}'}$ with the representation $\mathcal{S}$
\begin{equation}\label{equ.V-rep-algebra}
    \mathcal{M}_{\mathfrak{D}}\mathcal{M}_{\mathfrak{D}'} = \varphi(\mathfrak{D},\mathfrak{D}')\mathcal{M}_{\mathfrak{D}\ominus\mathfrak{D}'}.
\end{equation}

A similar picture also holds for the RDM. we can  define a set of $\mathcal{S}$-operators
\begin{equation}\label{equ.U-A-def}
    \mathcal{S}^{(A)}_\mathfrak{D} \equiv \hat{\Pi}_A\Sigma_\mathfrak{D}
\end{equation}
supported on the subregion $A$. With these newly defined operators, the RDM becomes
\begin{equation}\label{equ.red-DM-new}
    \rho_A = \frac{1}{2^{N^A-N^A_p}}\sum_{[\mathfrak{D}]\subset A}\mathfrak{C}_{\mathfrak{D}}\mathcal{S}^{(A)}_\mathfrak{D}.
\end{equation}
The form of the RDM above implies that the physical picture of Majorana fermions emerging at the endpoints of string operators still holds in any simply-connected subregion $A$, and it can be verified that $\mathcal{S}^{(A)}$-operators also satisfy a relation
\begin{equation}\label{equ.U-A-rep-algebra}
    \mathcal{S}^{(A)}_{\mathfrak{D}}\mathcal{S}^{(A)}_{\mathfrak{D}'} = \varphi(\mathfrak{D},\mathfrak{D}')\mathcal{S}^{(A)}_{\mathfrak{D}\ominus\mathfrak{D}'},
\end{equation}
which is the same as Eq.~\eqref{equ.U-rep-algebra} and \eqref{equ.V-rep-algebra}. The RDM of Majorana fermions in the region $A$ also possesses a similar form 
\begin{equation}\label{equ.rho-F-reduced}
    \rho_A^{F} \equiv \frac{1}{2^{N^A/2}}\sum_{[\mathfrak{D}]\subset A}\mathfrak{C}_{\mathfrak{D}}\mathcal{M}_\mathfrak{D},
\end{equation}
which is constructed similarly to Eq.~\eqref{equ.rho-F}. Here, we consider only regions with an even $N^A$, since the denominator $2^{N^A/2}$ becomes fractional if $N^A$ is odd, rendering the Hilbert space dimension ill-defined. In the latter sections, cases with an odd number of $N^A$ will be discussed. 

In conclusion, the comparison between the exact form of the DM and RDM in different representations demonstrates that the Majorana fermion representation is identical to the spin representation, and the Majorana fermions can be seen as emerging from the endpoints of those Pauli strings defined by Eq.~\eqref{equ.string-operator-def}~\cite{Kitaev_2006, fermion-and-string, string-net, Photons-and-electrons-as-emergent-phenomena-string, PRD-string-wen-quantum-order}. The identity between the spin representation and the Majorana-fermion representation arises from the fact that every operator $\mathcal{S}_\mathfrak{D}$ depends only on the equivalence class $[\mathfrak{D}]$ the configuration $\mathfrak{D}$ belongs to. In other words, the following operators with different configurations, which share the same set of endpoints, are identical.
\[
\mathcal{S}\lr{\wordfig{0.5}{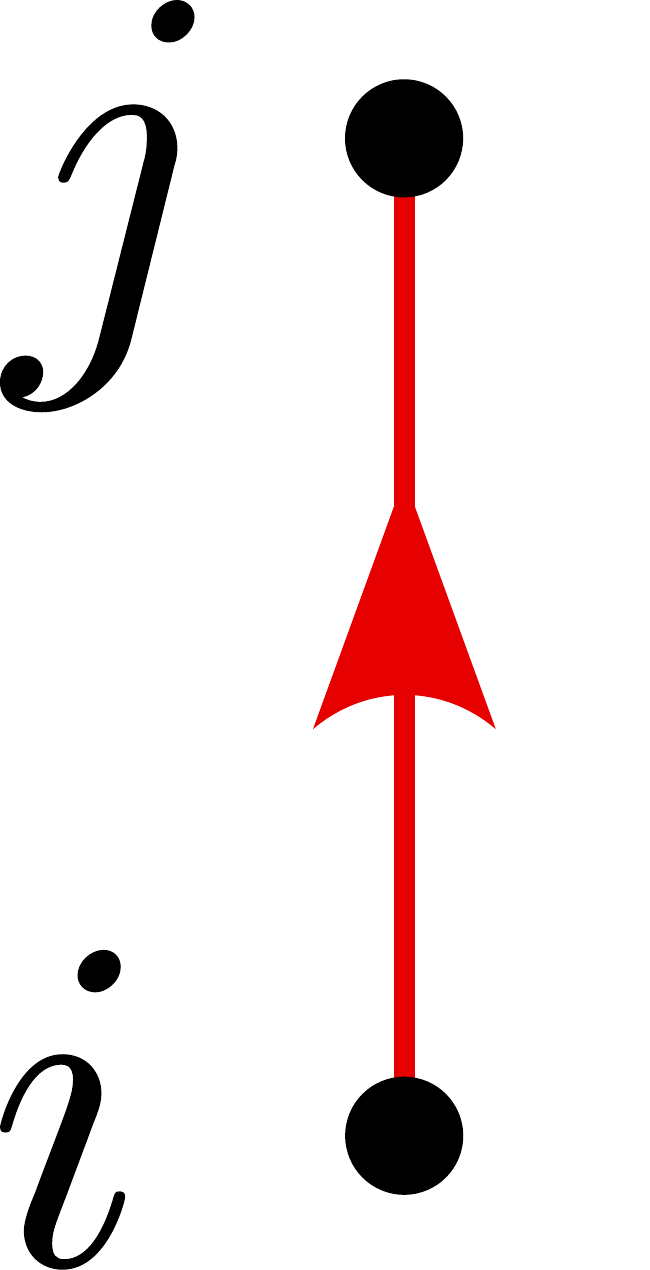}} = \mathcal{S}\lr{\wordfig{0.75}{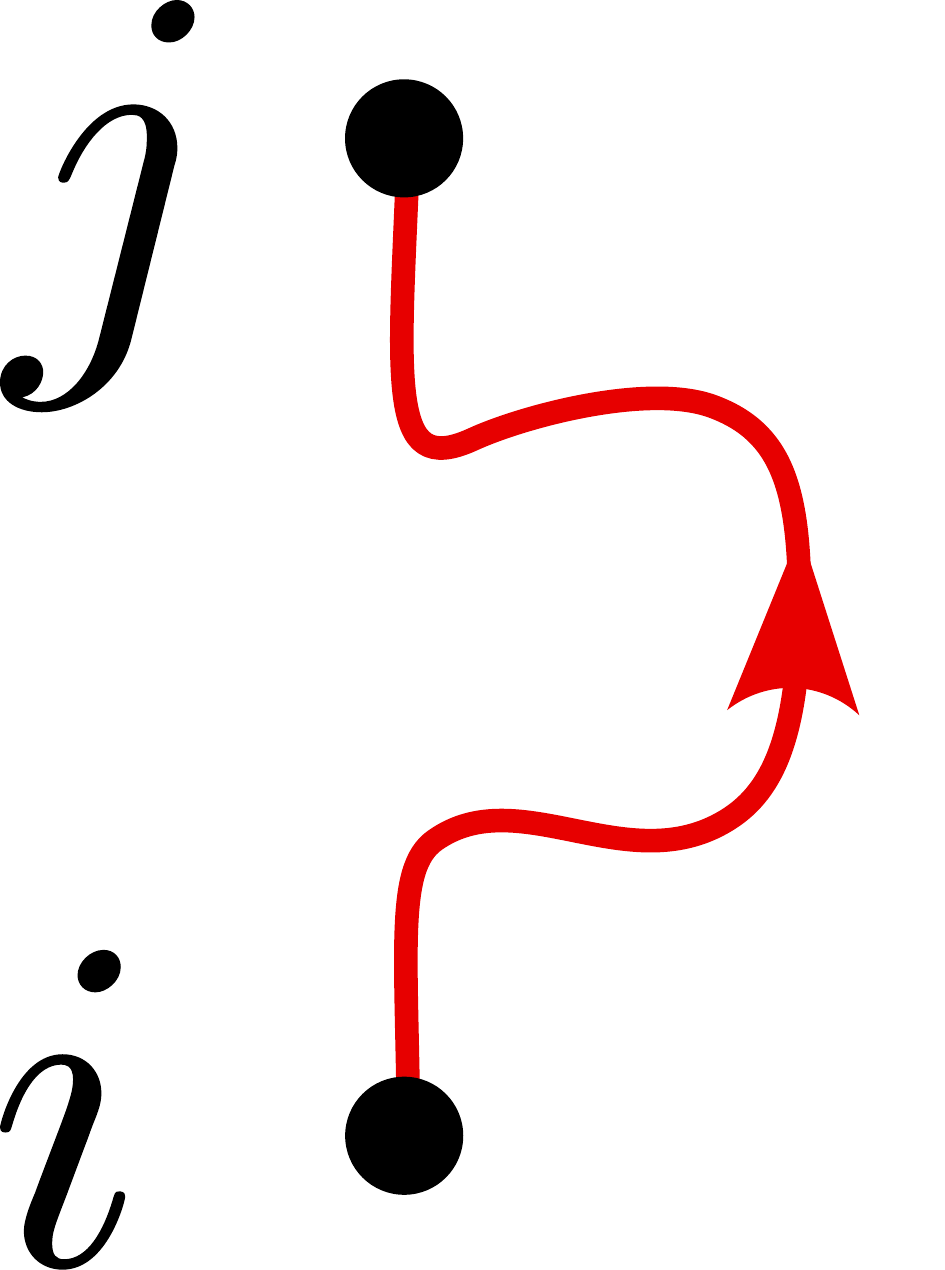}} = \mathcal{S}\lr{\wordfig{0.8}{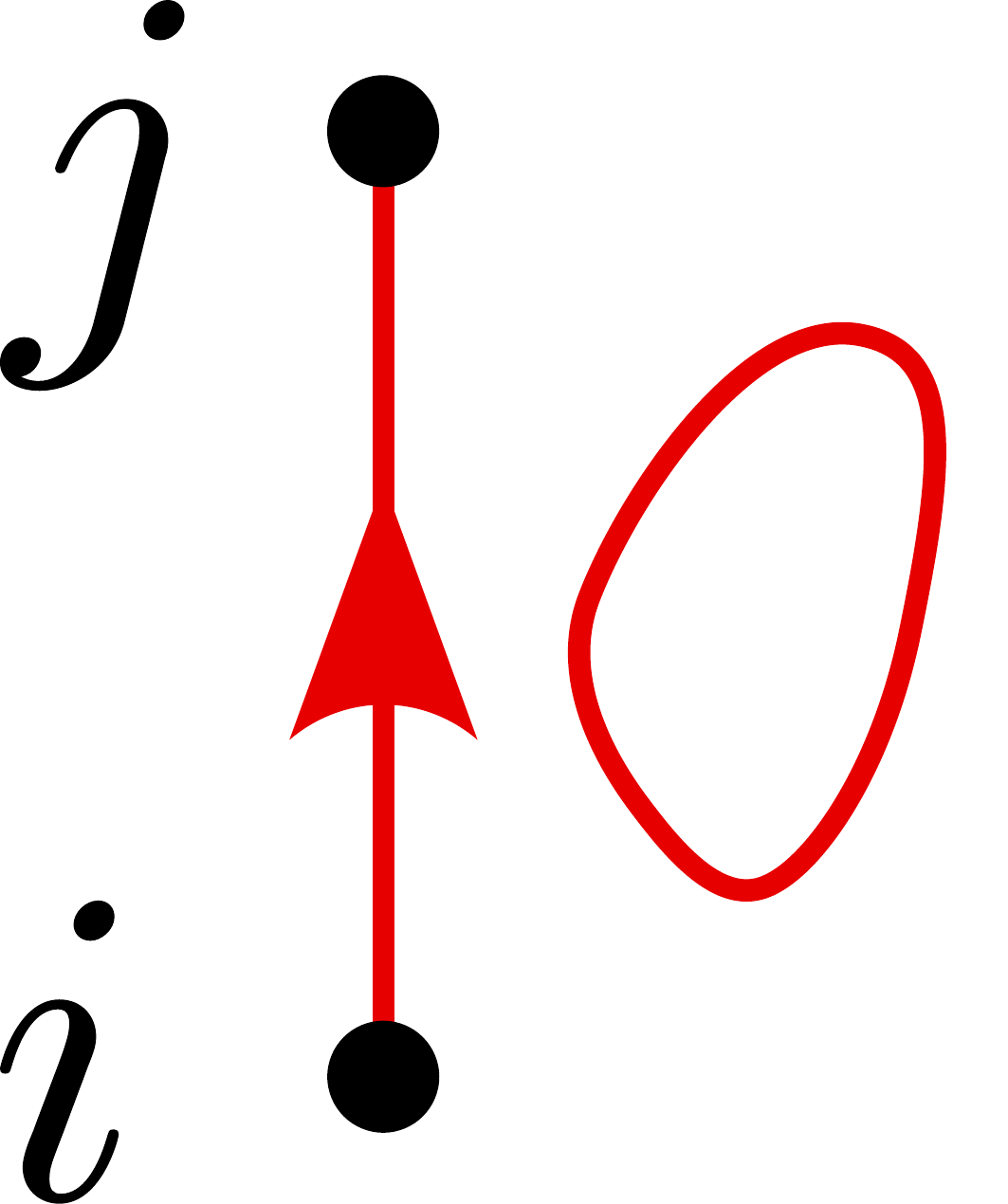}} = \cdots
\]
From the property of string operators $\Sigma_\mathfrak{D}$, we can directly prove the following equation
\begin{equation}\label{eq.S-rep-swap}
    \lsegment{i}{l}\lsegment{k}{ij} = -\lsegment{i}{j}\lsegment{k}{il},
\end{equation}
which can be represented pictorially as follows
\[
\mathcal{S}\lr{\wordfig{1.5}{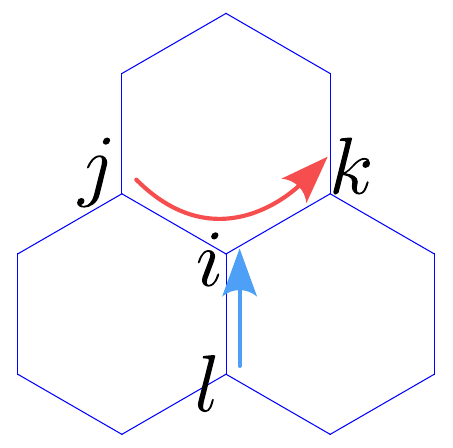}} = -\mathcal{S}\lr{\wordfig{1.5}{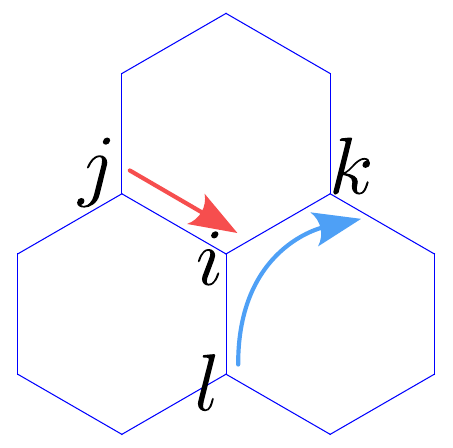}}.
\]
The physical meaning of this equation is that the endpoints of strings obey fermion statistics~\cite{Kitaev_2006, fermion-and-string, Kawagoe-Levin_2020, Liu-Lam-Ma-Zou_2024}. 
In fact, if we identify the operator $\lsegment{j}{i}$ of a single string $\overleftarrow{ji}$ pointing from $i$ to $j$ with the hopping operator $ic_jc_i$, which is actually the Majorana representation of the string $\mathcal{M}_{\overleftarrow{ji}}\propto ic_jc_i$, of Majorana fermions hopping from $i$ to $j$, then the above equation is equivalent to the following Majorana representation
\[
(ic_ic_l)(ic_kc_i)(ic_ic_j) = - (ic_ic_j)(ic_kc_i)(ic_ic_l),
\]
or equivalently
\begin{equation}\label{eq.M-rep_swap}
    \mathcal{M}_{\overleftarrow{il}} \mathcal{M}_{\overleftarrow{kij}}
    =
    -\mathcal{M}_{\overleftarrow{ij}}
    \mathcal{M}_{\overleftarrow{kil}}.
\end{equation}
Comparing Eq.~\eqref{eq.M-rep_swap} with Eq.~\eqref{eq.S-rep-swap} implies the identity between the $\mathcal{S}$-representation and the Majorana representation.

\section{Entanglement Structures}\label{sec:entanglement-structures}
In this section, we discuss the entanglement structure of the Kitaev ground state, including the EE and the ES, by directly analyzing the RDM we obtained in previous sections. 

We address three major issues. First, our method for calculating the EE, elucidated in Sec.~\ref{sec:EE-calculation-even}, improves on traditional methods that rely on formulating the theory in terms of fermions and gauge fields, where the issue of cutting gauge degrees of freedom living on the edges arises. Second, in Sec.~\ref{sec:EE-calculation-odd}, we explain how our approach overcomes the limitation of the replica trick, in which the analysis is well-defined only if the subregion has an even number of sites~\cite{Yao-Qi_2010}, by observing a relation between the subsystem fermion parity and the entanglement, which also gives hints about the entanglement structure of cases with an even number of sites. Lastly, we formulate everything in the spin representation to illustrate the microscopic origin of the entanglement structure through an analysis of the ES. Our approach is based on the theory proposed by \cite{Xu-Rakovszky-Knap-Pollmann_2025}. We begin by manifesting the block-diagonal structure of the DM in Sec.~\ref{sec:block-diagonal-structure}, which gives a massive degeneracy in the ES. In Sec.~\ref{sec:origin_of_degeneracy}, we further discuss the physical meaning of the degeneracy and its relation to the separability of the entanglement in the spin representation. Finally, in Sec.~\ref{sec:origin_degeneracy_odd}, we illustrate the physical meaning of the interplay between the entanglement and the subsystem fermion parity in the spin representation, indicating the radical difference between cases with an odd and even number of sites.

\subsection{Entanglement Entropy and its Separability}\label{sec:EE-calculation-even}

In this section, the EE of the Kitaev ground state is calculated using the DM and the algebras of strings obtained earlier, which reproduces the result proposed by~\cite{Yao-Qi_2010} that it can be separated into the gauge and the matter contribution. The derivation reveals that the matter part of the EE reflects the equivalence between representations $\mathcal{S}_\mathfrak{D}$ and $\mathcal{M}_\mathfrak{D}$, while the gauge part arises from the difference in dimension of the Hilbert space between these two identical representations.

Due to the equivalence of the representations $\mathcal{S}^{(A)}$ and $\mathcal{M}$ in the region $A$, plus the similarity of the forms of Eq.~\eqref{equ.red-DM-new} and \eqref{equ.rho-F-reduced}, the EE of the spin system ${S_A = -\Tr_A\rho_A\ln\rho_A}$ can be calculated using the Majorana DM \eqref{equ.rho-F-reduced} as an alternative. Inspired by this insight, we first observe that both representations $\mathcal{S}^{(A)}$ and $\mathcal{M}$ consist of traceless operators and the identity. Therefore, their trace satisfies 
\begin{equation}\label{equ.U-V-trace-relation}
    \left\{
\begin{aligned}
&\Tr_{A} \mathcal{S}^{(A)}_{\mathfrak{D}} = 2^{N^A-N^A_p}\delta_{\mathfrak{D},\varnothing} \\
& \Tr_{A,F} \mathcal{M}_{\mathfrak{D}} = 2^{N^A/2}\delta_{\mathfrak{D},\varnothing}
\end{aligned}
\right.
\end{equation}
where $\Tr_{A,F}$ stands for taking the trace over the matter sector in the region $A$. For simplicity, we assume the subregion $A$ contains an even number of lattice sites $N^A$. Cases with an odd number of lattice sites will be discussed in later sections. Given that the two representations are identical, the following identity holds for an arbitrary analytic function $f$: 
\begin{equation}\label{equ.identity}
    \Tr_A f(\{\mathcal{S}^{(A)}\}) = 2^{N^A/2 - N^A_p} \Tr_{A,F} f(\lrg{\mathcal{M}}).
\end{equation}
Given this property, we can define a function
\begin{align*}
    &f(\lrg{\mathcal{R}}) \\
    \equiv& \lrm{\frac{1}{2^{N^A-N^A_p}}\sum_{[\mathfrak{D}]\subset A}\mathfrak{C}_\mathfrak{D}\mathcal{R}_{\mathfrak{D}}}\ln\lrm{\frac{1}{2^{N^A-N^A_p}}\sum_{[\mathfrak{D}]\subset A}\mathfrak{C}_\mathfrak{D}\mathcal{R}_{\mathfrak{D}}},
\end{align*}
where $\mathcal{R}$ stands for an arbitrary representation of the algebra $\anglelr{\lrg{\mathfrak{D}},\ominus}$. With this,{the following two functions that appear in the formula for computing the entropy can be expressed in terms of the function $f$ with different representations as the domain}.
\begin{equation}\label{eq:rholnrho}
\left\{
\begin{aligned}
\rho^F_A\ln\rho_A^F &= 2^{N^A/2-N^A_p} f\lr{\lrg{\mathcal{M}}}  
    + \lrm{\lr{\frac{N^A}{2}-N^A_p}\ln 2}\rho^F_A
    \\
\rho_A\ln\rho_A &= f(\{\mathcal{S}^{(A)}\})
\end{aligned}
\right.
\end{equation}
The EE can be written in terms of the above functions, and the trace over the spin space can be converted to the trace over the matter sector using Eq.~\eqref{equ.identity}
\begin{align*}
    S_A &= -\Tr_A f(\{\mathcal{S}^{(A)}\}) \\
    &= -2^{N^A/2 - N^A_p}\Tr_{A,F} f(\lrg{\mathcal{M}}) \\
    &= -\Tr_{A,F} \rho^F_A\ln\rho^F_A + \lr{\frac{N^A}{2}-N^A_p}\ln 2.
\end{align*}
{The last term proportional to $\ln 2$ can be converted to the area-law entanglement.} To demonstrate this, we introduce a geometric relation connecting $N^A$, $N^A_p$, and $\abs{\de A}$ (the proof is given in Appendix~\ref{appendix:Pick}),
\begin{equation}\label{equ.Pick_sim-con}
    \abs{\de A} = N^A - 2N^A_p + 2.
\end{equation}
Due to this relation, the term proportional to $\ln 2$ in the last line of the above derivation can be rewritten in terms of $\abs{\de A}$, 
\begin{equation}\label{equ.EE-separation}
    S_A = \underbrace{-\Tr_{A,F}\lr{\rho_A^F\ln\rho_A^F}}_{\equiv S_A^F} + \underbrace{\lr{\frac{1}{2}\abs{\de A}-1}\ln 2}_{\equiv S_A^G}.
\end{equation}
This formula shows that the EE can be separated into the matter part $S^F_A$ and the gauge part $S^G_A$, in agreement with the result in Ref.~\cite{Yao-Qi_2010}. 
{
The decomposition in Eq.~\eqref{equ.EE-separation} has a direct structural interpretation at the level of the physical RDM. 
As we demonstrate in the following sections, the matter contribution is encoded in the eigenvalue spectrum within each identical block of the RDM, whereas the gauge contribution originates from the exact multiplicity of these blocks and therefore from the associated degeneracy of the entanglement spectrum.
This block organization is enforced by the gauge structure of the physical spin Hilbert space, which becomes explicit through the Gauss-law constraint and the exact 1-form Wilson symmetry introduced in sections following Sec.~\ref{sec:block-diagonal-structure}.   
Thus, the gauge--matter separation of the EE is not merely an entropy-level identity, but reflects a more fundamental organization of the physical RDM.
Below, we discuss the two contributions separately.
}

{
The matter contribution to the EE calculated from Eq.~\eqref{equ.rho-F-reduced} depends on the model details encoded in the spin correlation functions. It can be computed exactly in the Kitaev model because the equivalent Majorana fermion system is free, where all information about the state is encoded in the two-point correlation functions~\cite{Jin-Korepin_2004, Vidal-Latorre-Rico-Kitaev_2003, Latorre-Rico-Vidal_2004, Peschel_2003, Peschel-Eisler_2009, Chung-Peschel_2001, Cheong-Henley_2004}. The single-particle ES is the eigenvalue spectrum of the correlation matrix $\Gamma^{(c)}$, which is of size $N^A\times N^A$, defined as
\begin{equation}
    [\Gamma^{(c)}]_{ij} \equiv \frac{1}{2}\anglelr{c_ic_j}\quad ;\quad i,j\in A.
\end{equation}
The eigenvalues of $\Gamma^{(c)}$, denoted by $\gamma_{k,\pm}$, can be expressed in pairs if $N^A$ is even~\cite{Jin-Korepin_2004, Latorre-Rico-Vidal_2004}
\begin{equation}\label{equ.gamma-eigenvalue}
    \gamma_{k,\pm} = \frac{1 \pm \nu_k}{2} \quad ; \quad k = 1,2,\cdots, \frac{N^A}{2}.
\end{equation}
In this way, the contribution of the matter to EE becomes 
\begin{equation}\label{equ.EE-corr-mat}
    S_A^F = -\tr\lr{\Gamma^{(c)}\ln\Gamma^{(c)}}.
\end{equation}
The trace ``$\tr$'' is written in lowercase here to distinguish it from the trace ``$\Tr$'' over the Hilbert space of the quantum many-body system. This quantity can be computed via numerically diagonalizing the correlation matrix $\Gamma^{(c)}$. In terms of the correlation matrix, Eq.~\eqref{equ.EE-separation} can be written as
\begin{equation}\label{equ.Yao-Qi-standard-form}
    S_A = -\tr\lr{\Gamma^{(c)} \ln \Gamma^{(c)}} + \lr{\frac{1}{2}\abs{\de A} - 1}\ln 2.
\end{equation}

The gauge contribution, by contrast, originates from the mismatch in dimension between the physical spin Hilbert space and the equivalent Majorana-fermion Hilbert space justified by the algebraic equivalence between the spin representation $\mathcal{S}$ and the Majorana representation $\mathcal{M}$ of the Kitaev ground state, as represented by the term proportional to $\ln 2$ appearing in Eq.~\eqref{eq:rholnrho}. This holds independently of specific model details such as coupling constants or lattice geometry. This Hilbert-space dimensional mismatch, combined with the geometric relation~\eqref{equ.Pick_sim-con}, naturally gives rise to the area-law scaling of $S^G_A$. Here, the scaling coefficient $1/2$ indicates that each gauge degree of freedom residing on the edges crossing the entanglement cut contributes only half a bit of entropy, as the other half is accounted for by the fractionalized matter fermions. At this stage, the topological entanglement entropy $-\ln 2$ appearing in $S^G_A$ originates from the geometric relation~\eqref{equ.Pick_sim-con}. We elucidate its physical interpretation and its connection to the boundary-variable parity in sections following Sec.~\ref{sec:block-diagonal-structure}. Unlike the matter contribution $S^F_A$, which depends on model details, the explicit derivation of $S^G_A$ indicates that the gauge contribution to the EE relies on a more fundamental property: namely, the block organization of the physical RDM, as mentioned earlier.
}

\subsection{Interplay between Entanglement and Subsystem Fermion Parity}\label{sec:EE-calculation-odd}

A subtlety arises when the number of lattice sites in a subregion is odd, since a well-defined Hilbert space of a Majorana fermion system must have an even number of Majorana fermions. We address the problem in this section and demonstrate how this even-odd problem, i.e., the \textit{fermion parity}, is intertwined with the entanglement structure. Understanding subsystem fermion parity provides crucial insights into formulating and resolving this ambiguity directly within the physical spin representation.

To resolve the issue of odd $N^A$, we add a redundant Majorana fermion, denoted by $c_{(N^A+1)}$, to the original system, making the number of Majorana fermions even, so that our previous approach works again~\cite{Lee-Wilczek_2013}. We design the redundant Majorana fermion in such a way that it has no interaction with other Majorana fermions. Mathematically, we set $\sanglelr{ c_{(N^A+1)}c_i}=0\;\forall i\in A$. In this way, the dimension of the Hilbert space of the matter sector, including the redundant one, is $2^{(N^A+1)/2}$, and the RDM becomes
\begin{equation}\label{equ.rho-A-F-odd}
    \rho_A^{F} = \frac{1}{2^{\lr{N^A+1}/2}}\sum_{[\mathfrak{D}]\in A}\mathfrak{C}_{\mathfrak{D}}\mathcal{M}_\mathfrak{D}
\end{equation}
The redundant Majorana fermion does not enter the DM. Thus, the form of $\rho^F_A$ remains intact. Only the normalization coefficient is changed. With the redundant Majorana fermion, the original correlation matrix $\Gamma^{(c)}$ is correspondingly enlarged by one dimension to $\Tilde{\Gamma}^{(c)}$. According to the definition of $c_{(N^A+1)}$, we obtain
\begin{equation}\label{equ.tilde-Gamma-corner}
    \Tilde{\Gamma}^{(c)} = 
\lr{
\begin{array}{c|c}
    \Gamma^{(c)} & 0 \\
    \hline
    0 & 1/2
\end{array}
}.
\end{equation}
In the enlarged correlation matrix $\Tilde{\Gamma}^{(c)}$, the upper-left block is the original correlation matrix. The upper-right column and the lower-left row are zeros because the redundant Majorana fermion does not correlate with others. The element $1/2$ at the lower-right entry is given by the redundant Majorana fermion $\frac{1}{2}=\frac{1}{2}\sanglelr{c_{(N^A+1)}c_{(N^A+1)}}$. 

Due to the change in the dimension of the Hilbert space of the matter sector, the relation \eqref{equ.U-V-trace-relation} needs to be modified to the following
\begin{equation}\label{equ.U-V-trace-relation-odd}
    \left\{
\begin{aligned}
&\Tr_{A} \mathcal{S}^{(A)}_{\mathfrak{D}} = 2^{N^A-N^A_p}\delta_{\mathfrak{D},\varnothing} \\
& \Tr_{A,F} \mathcal{M}_{\mathfrak{D}} = 2^{(N^A+1)/2}\delta_{\mathfrak{D},\varnothing}
\end{aligned}
\right..
\end{equation}
Performing the same calculation as we did in the cases with even $N^A$, we obtain a result similar to the EE 
\begin{align}\label{equ.EE-odd-derive-1}
    S_A &= -\Tr_{A,F} \rho^F_A\ln\rho^F_A + \lr{\frac{N^A-1}{2}-N^A_p}\ln 2 \nonumber\\
    &= -\tr\lr{\Tilde{\Gamma}^{(c)}\ln \Tilde{\Gamma}^{(c)}} + \lr{\frac{1}{2}\abs{\de A} - \frac{3}{2}}\ln 2,
\end{align}
where the correlation matrix in Eq.~\eqref{equ.EE-corr-mat} is replaced by the enlarged one. According to Eq.~\eqref{equ.tilde-Gamma-corner}, the first term of Eq.~\eqref{equ.EE-odd-derive-1} can be reduced and expressed in terms of the original correlation matrix 
\begin{equation}\label{equ.EE-reduce}
    \tr\lr{\Tilde{\Gamma}^{(c)} \ln \Tilde{\Gamma}^{(c)}} = \tr\lr{\Gamma^{(c)} \ln \Gamma^{(c)}} - \frac{1}{2}\ln 2 .
\end{equation}
Substituting Eq.~\eqref{equ.EE-reduce} into Eq.~\eqref{equ.EE-odd-derive-1}, the last term $-\frac{1}{2}\ln 2$ in Eq.~\eqref{equ.EE-reduce} will cancel the term $-\frac{1}{2}\ln 2$ in Eq.~\eqref{equ.EE-odd-derive-1} arising from the enlarged dimension of the matter sector, and we ultimately obtain a result that is the same as Eq.~\eqref{equ.Yao-Qi-standard-form} for even $N^A$. 

The above derivation not only confirms that the result in Ref.~\cite{Yao-Qi_2010} remains valid for subsystems with an odd number of sites $N^A$, but also provides a new physical insight into the interplay between the matter and gauge sectors. Operationally, we first borrow a dimension $2^{1/2}$ of a single Majorana fermion from the gauge sector to make the Hilbert space of the matter sector well-defined. The resulting shift of $-\frac{1}{2}\ln 2$ reflects this borrowed dimension. Subsequently, the redundant fermion yields the term $+\frac{1}{2}\ln 2$, effectively restoring the dimension back to the gauge sector and leaving no net change. 

Having established the entropy formula for odd $N^A$, we now examine the ES in this situation and its intimate connection with the subsystem fermion parity. As shown in Eq.~\eqref{equ.gamma-eigenvalue}, the elements in the single-particle ES come in pairs. Since the lower-right corner of Eq.~\eqref{equ.tilde-Gamma-corner} provides an eigenvalue $1/2$, there must be another identical eigenvalue $\frac{1}{2}=1-\frac{1}{2}$ in the original single-particle ES. Therefore, the many-body ES can be written as 
\begin{equation}\label{equ.many-body-entanglement-spectrum-odd}
\lambda_{n_1,n_2,\cdots,n_{(N^A+1)/2}} = \frac{1}{2}\prod_{k=1}^{\frac{N^A-1}{2}}\gamma_{k,n_k}
\end{equation}
The formula shows that the many-body ES depends on only $(N^A-1)/2$ quantum numbers, although the complete labeling contains $(N^A+1)/2$ of them. In other words, there is a twofold degeneracy in the ES of the matter sector. Thus, we only have $2^{\frac{N^A-1}{2}}$ distinct values of $\lambda_{\lrg{n}}$. The degree of degeneracy for each distinct eigenvalue is therefore $2^{\frac{N^A+1}{2}-N^A_p}$ because the total number of eigenvalues is equal to the number of distinct eigenvalues multiplied by the number of degeneracies
\[
2^{N^A-N^A_p} = 2^{\frac{N^A-1}{2}}\times 2^{\frac{N^A+1}{2}-N^A_p}.
\]
Due to the conservation of fermion parity in the subregion $A$, the fermion parity operator, defined by $\hat{\bog{\pi}}^{(A)} = i^{N^{A}}\prod_{j\in A}c_j$, commutes with the RDM. From the fermion-parity point of view, the twofold degeneracy in the matter sector can be understood as a block-diagonalization that depends on the subsystem fermion parity. As we have seen, the value of $\lambda_{\lrg{n}}$ is independent of the last fermion number $n_{(N^A+1)/2}$. At the same time, $\hat{\bog{\pi}}^{(A)}$ commutes with the entanglement Hamiltonian, and hence also commutes with $\rho_A$. Therefore, ES divides into even and odd sectors, and these two identical sectors correspond to $\lrg{n}$ with $n_{(N^A+1)/2}=0,1$, respectively, giving a twofold degeneracy in ES of the matter sector. The microscopic origin of this degeneracy in terms of the physical spin RDM is discussed in Sec.~\ref{sec:origin_degeneracy_odd}.

\subsection{Block-Diagonal Structure of Density Matrices}\label{sec:block-diagonal-structure}

The fact that the spin representation $\mathcal{S}$ and the Majorana representation $\mathcal{M}$ are identical implies that the RDM might be of a block-diagonal form, in which the blocks are all identical. In this section, we discuss those operators that block-diagonalize the DM and elucidate their relation with the emergent gauge field. 

In light of the physical pictures given by \cite{Yao-Qi_2010} and \cite{Xu-Rakovszky-Knap-Pollmann_2025}, the operators that block-diagonalize the DM are defined at the boundary of the subregion because the block-diagonal structure of the RDM is due to the Gauss law of the gauge field~\cite{Xu-Rakovszky-Knap-Pollmann_2025}. Suppose the Gauss law in a gauge theory defined on a lattice is $\sum_{j\in \mathcal{N}_i}E_{ij} = \rho_i$, where $\rho_i$ is the matter field on the lattice site $i$, $E_{ij}$ is the electric field residing on the edge $\anglelr{ij}$, and $\mathcal{N}_i$ is the set of nearest neighbors of the site $i$. The boundary of the chosen subregion \textit{splits} the Gauss law of some neighborhoods $\mathcal{N}_i$ that cross the boundary, causing entanglement between the subregion and the environment because the Gauss law relates the matter field outside the subregion with the gauge field at the boundary, which is partly inside the subregion. Each \textit{{residual}} Gauss law at the boundary then acts as a symmetry operator that block-diagonalizes the RDM. 

In the Kitaev model, a problem is that the Gauss law of the gauge theory, which reads $b^x_ib^y_ib^z_ic_i = \mathds{1}$, only exists in the Majorana representation. Another issue is that the fundamental degrees of freedom of the Kitaev model reside on the lattice sites of the lattice, but the electric field operators $E_{ij}$ of the gauge theory, defined by $\mrm{e}^{i E_{ij}} = b^{\alpha_{ij}}_i$, live on edges, and which are not well-defined in the original setting of the model. For these reasons, we cannot directly apply the method proposed by~\cite{Xu-Rakovszky-Knap-Pollmann_2025} to a gauge theory that emerges from the parton construction. 

Our solution to this issue is to transform a Majorana fermion theory coupled to a gauge field back to the physical spin representation by \textit{gluing} pairs of fermions using string operators. Suppose there are an even number $\abs{\de A} = 2L$ of edges that cross the boundary of the simply-connected region $A$. The corresponding set of $b$-Majorana fermions encoding the residual electric field is $\lrg{b^{\mu_1}_{i_1}, b^{\mu_2}_{i_2}, b^{\mu_3}_{i_3},\cdots,b^{\mu_{2L}}_{i_{2L}}}$, where $\mu_k$ represents the direction of the $k$-th edge crossing the boundary, which is connected to site $i_k$. 
To transform these $b$-Majorana fermions to a physical representation, we insert gauge-field operators along strings $\mathcal{C}_1, \mathcal{C}_2, \mathcal{C}_3,\cdots, \mathcal{C}_{L}$ that link pairs of Majorana fermions $(b^{\mu_1}_{i_1},b^{\mu_2}_{i_2}), (b^{\mu_3}_{i_3},b^{\mu_4}_{i_4}),\cdots,(b^{\mu_{2L-1}}_{i_{2L-1}},b^{\mu_{2L}}_{i_{2L}})$, respectively, to create a series of physical operators of the form
\begin{equation}\label{equ.b-link}
    b^{\mu_{2n-1}}_{i_{2n-1}} \lr{\prod_{\anglelr{ij}\in\mathcal{C}_{n}} \hat{u}_{ij}}b^{\mu_{2n}}_{i_{2n}} \quad ; \quad n=1,2,3,\cdots, L
\end{equation}
in which the two $b$-Majorana fermions $b^{\mu_{2n-1}}_{i_{2n-1}}$ and $b^{\mu_{2n}}_{i_{2n}}$ are at the endpoints of the string $\mathcal{C}_n$, and the product of gauge-field operators in between is the string that links the two endpoints together. 
\begin{figure}
    \centering
    \includegraphics[width=\linewidth]{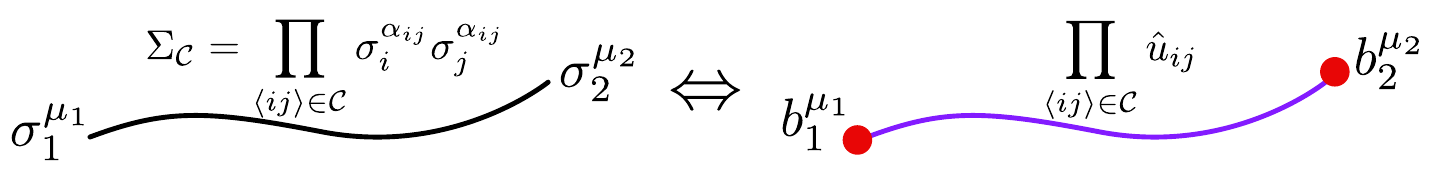}
    \caption{Two $b$-Majorana fermions can also be linked by a series of gauge-field operators, $\prod_{\anglelr{ij}\in\mathcal{C}} \hat{u}_{ij}$. In that way, a physical operator $\hat{\xi}$ is constructed.}
    \label{fig.linking-strings-electric-field}
\end{figure}
However, the expression \eqref{equ.b-link} only contains $b$-Majorana fermions and cannot be transformed directly to the spin representation. To obtain physical strings containing purely spin operators, we have to insert $c$-Majorana fermions into the expression \eqref{equ.b-link}, then we obtain the following result   
\begin{equation}\label{equ.c-b-link}
    b^{\mu_{2n-1}}_{i_{2n-1}}c_{i_{2n-1}} \lr{\prod_{\anglelr{ij}\in\mathcal{C}_{n}} c_i\hat{u}_{ij}c_j}c_{i_{2n}}b^{\mu_{2n}}_{i_{2n}}.
\end{equation}
After considering the transformation $\sigma^\mu_i\to ib^\mu_ic_i$, the operator \eqref{equ.c-b-link} is equal to the operators $\hat{\xi}_n$ defined in the following way
\begin{equation}\label{equ.spin-string-linking-gauge}
    \hat{\xi}_n \equiv \sigma^{\mu_{2n-1}}_{i_{2n-1}}\Sigma_{\mathcal{C}_n}\sigma^{\mu_{2n}}_{i_{2n}} \quad ; \quad n=1,2,3,\cdots ,L
\end{equation}
up to a phase that depends on the lengths of the strings $\abs{\mathcal{C}_n}$. The process of producing $\hat{\xi}$-operators from \eqref{equ.b-link} to \eqref{equ.spin-string-linking-gauge} is illustrated in \autoref{fig.linking-strings-electric-field}. All $\hat{\xi}_n$ operators defined in a simply-connected subregion $A$ commute with the RDM $\rho_A$. 
\begin{figure}
    \centering
    \includegraphics[width=0.7\linewidth]{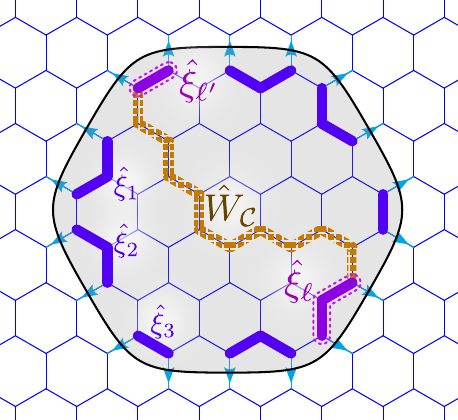}
    \caption{The $\hat{\xi}$-operators with their two endpoints attached to the boundary are defined by Eq.~\eqref{equ.spin-string-linking-gauge}, and the directions $\mu_{n}$ in the definition are the same as the residual electric field represented by the arrows crossing the boundary. A non-contractible Wilson operator $\hat{W}_{\mathcal{C}}$ given by Eq.~\eqref{eq.regional-global-wilson-def} is defined in the same way as the $\hat{\xi}$-operators. In the illustration, the two endpoints of the non-contractible Wilson operator are connected to $\hat{\xi}_\ell$ and $\hat{\xi}_{\ell'}$, respectively. Because of relation~\eqref{eq.wilson-xi-flip}, the signs of the two $\hat{\xi}$-operators are flipped by the non-contractible Wilson operator. } 
    \label{fig.xi-def}
\end{figure}
\autoref{fig.xi-def} illustrates the $\hat{\xi}$-operators defined at the boundary of the subregion using the method given by Eq.~\eqref{equ.spin-string-linking-gauge}. By exhausting all possible string operators in the subregion and the commutation relations among spin operators, we prove that $[\hat{\xi}_n,\rho_A]=0\;\forall n$ because $\rho_A$ consists of string operators inside $A$. At the same time, the regions where $\hat{\xi}$-operators are supported have no intersection, hence they commute with each other, $[\hat{\xi}_m,\hat{\xi}_n]=0\;\forall m,n$. Therefore, the set $\lrg{\xi_1,\cdots,\xi_L}$ of expectation values of $\hat{\xi}$-operators defined in the subregion $A$ forms good quantum numbers that block-diagonalizes the RDM
\begin{equation}\label{equ.rho-block-diagonal-structure}
    \rho_A = \bigoplus_{\lrg{\xi_1,\cdots,\xi_L}} \varrho_{\lrg{\xi_1,\cdots,\xi_L}}
\end{equation}
where each $\varrho_{\lrg{\xi}}$ is the block labeled by a set of binaries $\lrg{\xi}$, and thus the number of blocks is $2^{L} = 2^{\abs{\de A}/2}$.

\subsection{Origin of Degeneracy}\label{sec:origin_of_degeneracy}

The block-diagonal structure of $\rho_A$ implies a massive degeneracy in the ES, since all the blocks in the DM are identical, which will be proved in this section. Furthermore, we interpret the gauge part of the entanglement as corresponding to the degeneracy in the ES. We also explain the emergence of the matter part of the entanglement.

The $\hat{\xi}$-operators defined in Sec.~\ref{sec:block-diagonal-structure} block-diagonalize the RDM, with each block labeled by a binary string $\lrg{\xi}$ inspired by \cite{Xu-Rakovszky-Knap-Pollmann_2025}. {Proving that all blocks are identical implies that the RDM comprises $2^{L-1}$ identical copies, leading to a $2^{L-1}$-fold degeneracy in the ES.} To prove this, we have to construct unitary transforms that relate different blocks and commute with the DM. In a $\mathbb{Z}_2$ gauge theory with an exact 1-form Wilson symmetry, the operators that relate all the blocks are the non-contractible Wilson operators $\hat{W}_{\mathcal{C}}$, defined in the same way as the $\hat{\xi}$-operators, 
\begin{equation}\label{eq.regional-global-wilson-def}
    \hat{W}_\mathcal{C} \equiv \sigma^{\mu_i}_{i}\Sigma_{\mathcal{C}}\sigma^{\mu_f}_{f}.
\end{equation}
An example is illustrated in \autoref{fig.xi-def}. The two endpoints $i,f$ of the path $\mathcal{C}$ are attached to the boundary of the subregion, and $\mu_{i,f}$ are the directions of the residual gauge field at sites $i$ and $f$ that cross the boundary, respectively. In \autoref{fig.xi-def}, if $\hat{W}_{\mathcal{C}}$ and $\hat{\xi}_{\ell}$ have a common endpoint, the Wilson operators would anti-commute with $\hat{\xi}_\ell$ and flip its sign:
\begin{equation}\label{eq.wilson-xi-flip}
    \hat{W}_\mathcal{C} \hat{\xi}_{\ell}\hat{W}_{\mathcal{C}}^\dagger = -\hat{\xi}_{\ell}.
\end{equation}
Therefore, the non-contractible Wilson operator defined in the subregion relates different blocks of $\rho_A$ in the following way,
\begin{equation}\label{equ.related-blocks}
    \hat{W}_{\mathcal{C}} 
    \varrho_{
            \{\cdots,
            \xi_{\ell},
            \cdots,
            \xi_{\ell'},
            \cdots\}
        }
    \hat{W}_{\mathcal{C}}^\dagger
    =
    \varrho_{
            \{\cdots,
            -\xi_{\ell},
            \cdots,
            -\xi_{\ell'},
            \cdots\}
        }.
\end{equation}

Based on the block-diagonal structure \eqref{equ.rho-block-diagonal-structure} of $\rho_A$ and the fact that non-contractible Wilson operators relate different blocks in a way given by Eq.~\eqref{equ.related-blocks}, we conclude that there are 
\[2^{L-1} = 2^{\abs{\de A}/2 - 1}\] 
identical blocks in $\rho_A$. The factor $2^{-1}$ arises from the fixed parity of quantum numbers, which can also be understood as a consequence of the exact 1-form Wilson symmetry~\cite{Xu-Rakovszky-Knap-Pollmann_2025}. Based on this block-diagonal structure, $\rho_A$ can be expressed as
\begin{equation}\label{equ.block-rho-final}
    \rho_A = \frac{1}{2^{L-1}}\bigoplus_{n=1}^{2^{L-1}}\varrho^{(n)}_A
\end{equation}
with all the blocks $\varrho_A^{(n)}$ identical. Substitution of such an expression into the definition of von Neumann entropy directly gives the EE, 
\begin{align}\label{equ.EE-from-blocks}
    S_A 
    &= -\Tr_A\rho_A\ln\rho_A \nonumber \\
    &= -\Tr_A \varrho_A\ln\varrho_A + \lr{\frac{1}{2}\abs{\de A} -  1}\ln 2.
\end{align}
Comparing Eq.~\eqref{equ.EE-from-blocks} with Eq.~\eqref{equ.EE-separation}, we find that each block $\Tilde{\rho}_A$ corresponds to the Majorana RDM $\rho_A^F$. At the same time, the gauge part $S^G_A$ measures the number of identical blocks in $\rho_A$, which is exactly the degree of degeneracy in the ES. This agrees with the gauge nature of such a contribution since the degeneracy 
is yielded by the residual Gauss law after the environment is traced out. Finally, the topological EE $-\ln 2$ is a consequence of the topological constraint that the total parity of quantum numbers, $\prod_{n=1}^L\xi_n$, is fixed, originating from the exact 1-form Wilson symmetry. 

\begin{figure}
    \centering
    \includegraphics[width=0.7\linewidth]{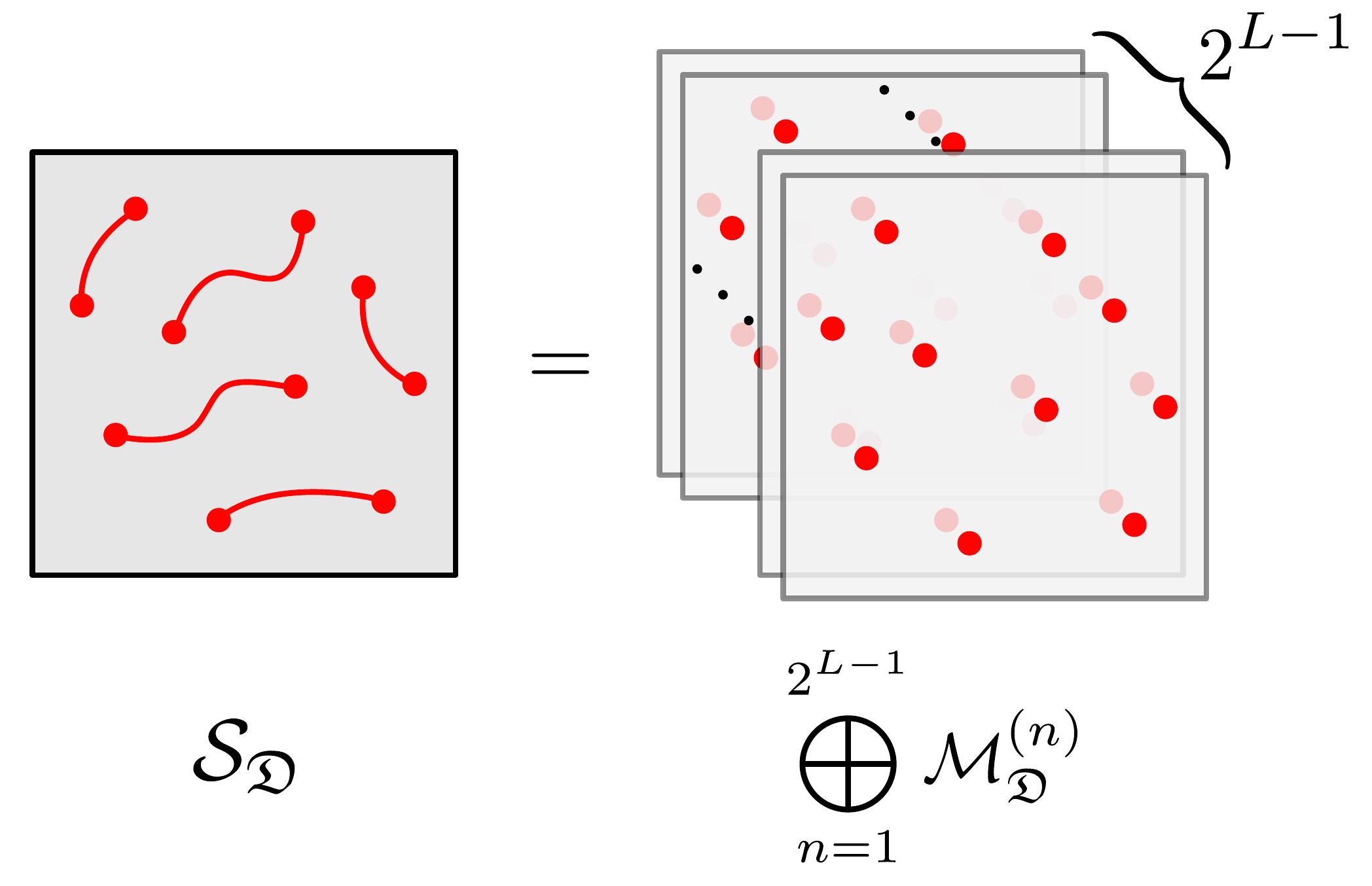}
    \caption{An illustration of the block-diagonal structure of the spin representation. An operator $\mathcal{S}_\mathfrak{D}$ can be seen as an assembly of $2^{L-1}$ copies of Majorana systems $\mathcal{M}_{\mathfrak{D}}$ with the identical set of endpoints $\mrm{EP}(\mathfrak{D})$. Therefore, an algebraic operation of $\mathcal{S}_\mathfrak{D}$ is equivalent to that of $2^{L-1}$ copies of the corresponding $\mathcal{M}_\mathfrak{D}$, which makes the spin representation identical to the Majorana fermion representation.}
    \label{fig:block-diag-illustration}
\end{figure}
From a more fundamental point of view, the set of $\hat{\xi}$-operators block-diagonalizes every string operator in a way identical to what we have for $\rho_A$ because each of them commutes with every string operator, hence so is $\mathcal{S}^{(A)}_{\mathfrak{D}}$
\[
[\hat{\xi}_n,\mathcal{S}^{(A)}_{\mathfrak{D}}]=0 \;\forall n\in\lrg{1,2,\cdots,L}
\]
Similarly to Eq.~\eqref{equ.block-rho-final}, every $\mathcal{S}$-operator can also be block-diagonalized as
\begin{equation}\label{equ.U-V-block-relation}
    \mathcal{S}^{(A)}_{\mathfrak{D}} = \frac{1}{2^{L-1}} \bigoplus_{n=1}^{2^{L-1}}\mathcal{M}^{(n)}_{\mathfrak{D}}
\end{equation}
where $\{\mathcal{M}^{(n)}_\mathfrak{D}\}$ is a set of identical blocks of $\mathcal{S}^{(A)}_{\mathfrak{D}}$. 
To ensure that expression \eqref{equ.U-V-block-relation} satisfies rule \eqref{equ.U-rep-algebra}, we also have to implement such an algebraic rule on every block $\mathcal{M}^{(n)}_{\mathfrak{D}}$, 
\[
\mathcal{M}^{(n)}_{\mathfrak{D}}\mathcal{M}^{(n)}_{\mathfrak{D}'} = \varphi(\mathfrak{D},\mathfrak{D}')\mathcal{M}^{(n)}_{\mathfrak{D}\ominus\mathfrak{D}'}
\] 
Based on this, we have confirmed that those blocks $\mathcal{M}^{(n)}_{\mathfrak{D}}$ in Eq.~\eqref{equ.U-V-block-relation} are precisely $\mathcal{M}$-operators defined in Eq.~\eqref{equ.V-rep-def}. In short, the gauge-field degree of freedom is actually the number of identical blocks of string operators represented by spins, and each identical block depicts the physics of free Majorana fermions. This precisely reflects the fractionalization of the spin degree of freedom into gauge-field and matter sectors, with the matter sector further fractionalizing into deconfined Majorana fermions.

\subsection{Interplay between Degeneracy and Subsystem Fermion Parity}\label{sec:origin_degeneracy_odd}

The fermion parity obtained in Sec.~\ref{sec:EE-calculation-odd} motivates a question: how do we describe the subsystem fermion parity in the spin representation? And how does it give an additional degeneracy in the ES if the subregion contains an odd number of lattice sites? In this section, we find a specific sort of operator in the spin representation corresponding to the subsystem fermion parity that further block-diagonalizes the DM if $N^A$ is odd, which gives an additional degeneracy in the EE.

\begin{figure}
    \centering
    \includegraphics[width=0.45\linewidth]{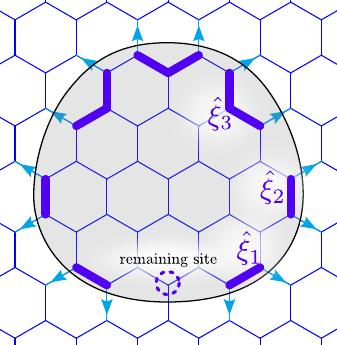}
    \hspace{0.5cm}
    \includegraphics[width=0.45\linewidth]{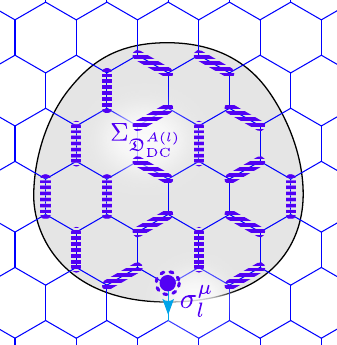}
    \caption{For cases with odd $N^A$, there must be a single site $l$ on the boundary that remains unpaired after defining $\hat{\xi}$-operators by pairing the adjacent edges crossing the boundary, as illustrated in the left panel. Based on this observation, we can define an operator $\hat{\Xi}_A\equiv \sigma^\mu_l \Sigma_{\mathfrak{D}_{\mrm{DC}}^{A(l)}}$ by placing a Pauli matrix $\sigma^\mu_l$ at the remaining site $l$ with $\mu$ the direction of the edge crossing the boundary that connects to $l$, and filling the rest of the part by a dimer covering $\Sigma_{\mathfrak{D}_{\mrm{DC}}^{A(l)}}$.}
    \label{fig.Xi-odd}
\end{figure}

A relation implied by Eq.~\eqref{equ.Pick_sim-con} 
\[%
    N^A=\abs{\de A} \mod{2},
\]
indicates that once the number of lattice sites in $A$ is odd, so is the length of the boundary. This raises an issue that there will always be one single site left after sorting lattice sites along the boundary in pairs, as shown in the left panel of \autoref{fig.Xi-odd}. If $L$ is defined as the number of $\hat{\xi}$-operators, then we only have
\[
L = \frac{1}{2}\lr{\abs{\de A} - 1},
\]
in cases with odd $\abs{\de A}$. In this case, the block-diagonal structure \eqref{equ.block-rho-final} and the derivation \eqref{equ.EE-from-blocks} are still valid, but the result should be modified as
\begin{align}\label{equ.EE-odd-blocks}
    S_A &= -\Tr_A\varrho_A\ln\varrho_A + \lr{L-1}\ln 2 \nonumber\\
    &= -\Tr_A\varrho_A\ln\varrho_A + \lr{\frac{1}{2}\abs{\de A}-\frac{3}{2}}\ln 2.
\end{align}
The form of Eq.~\eqref{equ.EE-odd-blocks} is the same as that of Eq.~\eqref{equ.EE-odd-derive-1}, showing that the block $\varrho_A$ here is identical to the Majorana RDM \eqref{equ.rho-A-F-odd} with a redundant Majorana fermion included. 

In Sec.~\ref{sec:EE-calculation-odd}, the result of the Majorana ES shows that the degree of degeneracy of a system with odd $N^A$ is 
\begin{equation}\label{eq:odd-degree-degeneracy}
    2^{\frac{N^A+1}{2}-N^A_p} = 2^{\lr{\abs{\de A} - 1}/2} = 2^L.
\end{equation}
However, as established previously for even site counts, a set of $L$ operators $\hat{\xi}$ yields a degeneracy of $2^{L-1}$. This discrepancy implies that apart from the set of $\hat{\xi}$-operators, there exists an additional operator that further block-diagonalizes $\rho_A$ to yield an additional twofold degeneracy in the ES. At the end of Sec.~\ref{sec:EE-calculation-odd}, we have elucidated that the additional twofold degeneracy in the ES of the Majorana sector originates from the subsystem fermion parity. Therefore, we anticipate that the additional operator that block-diagonalizes the DM, which will be denoted by $\hat{\Xi}_A$ later, should be related to the subsystem fermion parity in the Majorana representation. A feasible choice is a dimer covering of the subregion $A$,
\begin{equation}\label{equ.Xi-def}
    \hat{\Xi}_A \equiv \sigma^\mu_l \Sigma_{\mathfrak{D}_{\mrm{DC}}^{A(l)}},
\end{equation}
where the edge configuration $\mathfrak{D}_{\mrm{DC}}^{A(l)}$ stands for a dimer covering of $A$ with a site $l$ being excluded. Since $N^A$ is an odd number, it is not possible for a dimer configuration composed of non-overlapping dimers to cover the entire subregion~\footnote{The dimer covering is not uniquely defined. One can choose an arbitrary dimer covering with a remaining site $l$ at the boundary that covers the subregion.}. The remaining site $l$ is required to be at the boundary of the subregion $A$, and the direction $\mu$ of the Pauli matrix $\sigma_l^\mu$ is that of the edge connected to $l$ that crosses the boundary. See the right panel of \autoref{fig.Xi-odd}. 

$\hat{\Xi}_A$ can be examined to commute with $\rho_A$ in the same way as for $\hat{\xi}$-operators. String operators are composed of multiple dimers, and every dimer commutes with the dimer covering because of the properties of Pauli matrices. Therefore, the only thing to be concerned about is $\sigma^\mu_l$ lying at the remaining site.
As we have defined, the site $l$ is at the boundary and $\mu$ points outward. It can be seen directly that, inside $A$, there are only two edges connected to $l$, and the corresponding string operator of both edges anti-commutes with $\sigma^\mu_l$. However, since the string operator also anti-commutes with the dimer that connects to the other endpoint of the edge, the two contributions cancel each other. In the end, every string operator of $\mathfrak{D}\subset A$ commutes with $\hat{\Xi}_A$. Because $\rho_A$ is composed of string operators inside $A$, we conclude that $[\hat{\Xi}_A,\rho_A]=0$. 

$\hat{\Xi}_A$ is related to the subsystem fermion parity by transforming Eq.~\eqref{equ.Xi-def} into the Majorana fermion representation, and we find that it is equal to the subsystem fermion parity operator multiplied by the gauge-field operators arranged in order of the dimer covering $\mathfrak{D}^{A(l)}_{\mrm{DC}}$
\[
\hat{\Xi}_A \propto \lr{\prod_{\anglelr{ij}\in  \mathfrak{D}^{A(l)}_{\mrm{DC}}} \hat{u}_{ij}} \lr{b^\mu_l\prod_{i\in A}c_i}
\]
up to a phase. In the above formula, the subsystem fermion parity operator is defined as
\begin{equation}\label{equ.total-fermion-parity-odd}
    \hat{\bog{\pi}}^{(A)} \propto b^\mu_l\prod_{i\in A}c_i
\end{equation}
with a phase correction depending on the geometry of the subregion. Because the gauge has been fixed as $u_{ij}=1\;\forall\anglelr{ij}$, the expectation value of $\hat{\Xi}_A$ is completely determined by that of $\hat{\bog{\pi}}^{(A)}$. The definition of the subsystem fermion parity here is different from Eq.~\eqref{equ.total-fermion-parity-Sigma-Z}. Because $N^A$ is odd here, $\prod_{i\in A}c_i$ itself only contains an odd number of Majorana fermions, which is insufficient to define a fermion parity operator. This mechanism has been discussed in Sec.~\ref{sec:EE-calculation-odd}, and the alternative we adopted there is to include a redundant Majorana fermion $c_{(N^A+1)}$, and Eq.~\eqref{equ.total-fermion-parity-odd} implies that the redundant fermion is actually $b^\mu_l$ borrowed from the gauge sector. 

In summary, for cases with odd $N^A$, we have $L = \frac{1}{2}\lr{\abs{\de A} - 1}$ different $\hat{\xi}$-operators and an additional operator $\hat{\Xi}_A$ that commute with $\rho_A$. These operators give $2^{L+1}$ blocks labeled with a set of eigenvalues $\lrg{\xi_1,\cdots,\xi_L;\Xi_A}$. Here, the non-contractible Wilson operators still relate these blocks in the same way as we have discussed in the previous section. The only difference is that a non-contractible Wilson operator flips the sign of $\Xi_A$ only when one of its endpoints is at the remaining site $l$. {Together, there are $2^{L+1}/2 = 2^L$ identical blocks in the RDMs. Therefore, there is a $2^{L}$-fold degeneracy in the ES, in agreement with Eq.~\eqref{eq:odd-degree-degeneracy} and our analysis in Sec.~\ref{sec:EE-calculation-odd}.}

Although it seems that $\hat{\Xi}_A$ can also be defined in cases with even $N^A$ and leads to an additional twofold degeneracy, the operator $\hat{\Xi}_A$ defined in this way commutes with every non-contractible Wilson operator since the pattern is simply a dimer covering with no remaining site, implying that no operator can relate blocks $\Xi_A=+1$ and $\Xi_A=-1$.  Thus, $\hat{\Xi}_A$ cannot produce a twofold degeneracy in ES for even $N^A$. This highlights an intrinsic difference between systems with even and odd $N^A$, manifest as the subsystem fermion parity in the Majorana picture or, equivalently, as the dimer covering operator $\hat{\Xi}_A$ in the physical spin formulation.

\section{Summary and Discussions}\label{sec:summary}

In this work, we derive the explicit form of the DM of the Kitaev ground state on a torus and discuss the entanglement information encoded therein in detail. We use the spin formulation throughout this paper to address the lack of a theoretical understanding of quantum spin systems in the spin representation.

This work can be summarized in three main results. First, the ground state DM is composed of string operators in different equivalent classes of edge configurations. 
Regardless of the fermion parity condition, an equivalence class only depends on the pattern of endpoints of the edge configurations. This structure implies the equivalence between the spin representation and the fermion representation of edge configurations. 

Second, a RDM of a simply-connected subregion is obtained. The calculation of the EE implies that the RDM is composed of identical blocks, and each block is the DM of the corresponding free-fermion system. In other words, the ES has an extensive degeneracy, and the distinct values in the ES are the ones in the equivalent free-fermion system, and the gauge-field contribution to the EE measures the degeneracy in the ES. We also investigated the issue raised by cases with an odd number of lattice sites. Our approach solves this subtlety by borrowing a Majorana fermion from the gauge sector, and this derivation also yields {an additional twofold degeneracy in the EE of the fermion sector}.

Third, inspired by~\cite{Xu-Rakovszky-Knap-Pollmann_2025}, the block-diagonal structure of the RDM is directly proven. The physical picture is based on the \textit{residual} Gauss law of the gauge theory after the environment is traced out, and it then serves as real symmetries that give good quantum numbers for block-diagonalizing the RDM. After that, the blocks are proven to be identical because the exact 1-form Wilson symmetry relates all the blocks together. Since the gauge theory of the Kitaev model is only an emergent structure in the enlarged parton space, the fractionalized fermions need to be \textit{glued} back into physical spin degrees of freedom by inserting strings that connect pairs of Majorana fermions. Lastly, a global operator defined by dimer coverings that reflects the subsystem fermion parity gives an additional twofold degeneracy when the number of lattice sites is odd, in agreement with the twofold degeneracy found in the Majorana fermion sector.

\begin{figure}
    \centering
    \includegraphics[width=0.45\linewidth]{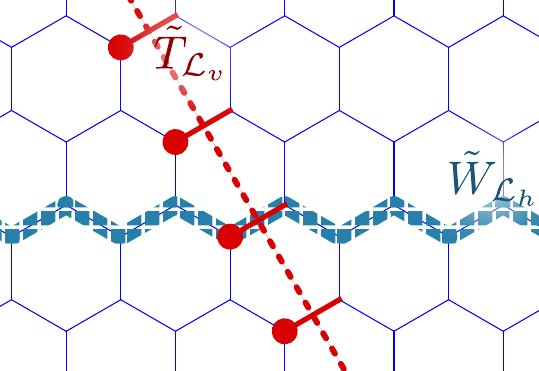}
    \hspace{0.2cm}
    \includegraphics[width=0.45\linewidth]{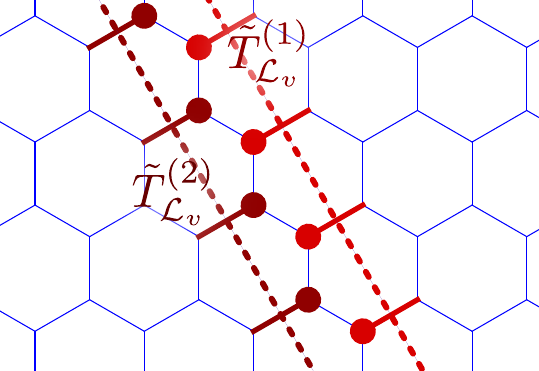}
    \caption{A 't~Hooft loop operator $\Tilde{T}_{\mathcal{L}}$ can be defined by multiplying Pauli matrices that flip the gauge field $u_{ij}$ along the closed path $\mathcal{L}$, which is equivalent to the path of a $\mathbb{Z}_2$ flux moving along the loop. In the left panel, a vertical non-contractible 't~Hooft loop defined by multiplying Pauli $x$-spins located at red dots $\Tilde{T}_{\mathcal{L}_v}=\prod_{i\in\mathcal{L}_v}\sigma^x_i$ anti-commutes with the horizontal non-contractible Wilson loop operator $\Tilde{W}_{\mathcal{L}_h} = \Sigma_{\mathcal{L}_{h}}$. However, the 't~Hooft loop operator defined in this way is not a symmetry operator since it does not commute with the Kitaev Hamiltonian. Moreover, it is not independent of the Wilson loop because two non-contractible 't~Hooft loop operators in the same direction combine to form a Wilson loop operator, as illustrated in the right panel.}
    \label{fig:t-hooft-loops}
\end{figure}

Although the discussion in Sec.~\ref{sec:entanglement-structures} of this work is mostly inspired by \cite{Xu-Rakovszky-Knap-Pollmann_2025}, there are still subtleties in certain aspects. It has been argued in \cite{Xu-Rakovszky-Knap-Pollmann_2025}, via a tensor-network approach, that the ground state of a system with an exact 1-form Wilson symmetry and a global $\mathbb{Z}_2$ symmetry must be at least twofold degenerate in the ES. However, in the Kitaev model, it is not clear whether such a contribution is included in the mechanism discussed in Sec.~\ref{sec:origin_of_degeneracy}. In addition, \cite{Xu-Rakovszky-Knap-Pollmann_2025} shows that the topological EE is a consequence of spontaneous 1-form 't~Hooft symmetry breaking. For the Kitaev model, however, even though the 1-form 't~Hooft loop operator can be understood as the path of a moving $\mathbb{Z}_2$ flux wrapping around one direction of the torus, as illustrated in Fig.~\ref{fig:t-hooft-loops}, we are not sure how to realize a real symmetry operator such that it commutes with the Hamiltonian and is independent of the Wilson loop operator simultaneously. We leave these problems to future investigations.

We would also like to mention that the form of the DM~\eqref{equ.density-matrix} is reminiscent of the tensor network state proposed by~\cite{Lee-Kaneko-Okubo-Kawashima_2019}, in which the tensor network \textit{Ans\"atze} that approximates the Kitaev ground state is constructed by applying a tensor product operator to an initial state. The tensor product operator is composed of a \textit{loop gas} operator and a \textit{dimer gas} operator, in which the loop gas term is exactly the plaquette projector~\eqref{eq.p-proj-def}, and the dimer gas part consists of low-order terms of the summation in Eq.~\eqref{equ.density-matrix}. The similarity between these two might be due to the projector nature of a pure state DM. If the tensor-product operator approximately captures the DM of the ground state, this might explain the remarkably good agreement between the results of \cite{Lee-Kaneko-Okubo-Kawashima_2019} and the exact solution.

{
Another remark is that the theoretical framework proposed in this work is general for every \textit{Kitaev-like} model as long as it is exactly solvable in a way similar to the Kitaev model, in which $\mathbb{Z}_2$ fluxes are static and the spin Hamiltonian can be completely separated into the gauge and matter parts~\cite{Lee-Zhang-Xiang_2007, yao-kivelson, Peri-Ok-Tsirkin-Neupert-Baskaran-Greiter-Moessner-Thomale_2020, Fu_2019}. 
In particular, some of these models host chiral spin liquid (CSL) ground states due to either the lattice geometry or higher-order interactions that break time-reversal symmetry~\footnote{
Regarding the lattice geometry, a plaquette operator could square to $-1$ if the plaquette is a polygon with an odd number of edges, implying time-reversal symmetry-broken states. See Eq.~\eqref{eq:sigma-joints} in Appendix~\ref{appendix:projective-rep} of the present work and Ref.~\cite{Peri-Ok-Tsirkin-Neupert-Baskaran-Greiter-Moessner-Thomale_2020} for details.}. Despite subtle differences in the plaquette projector, both string operators and equivalence classes remain valid, implying the same pattern of entanglement separability in these models and the same form of the gauge contribution to the EE.
}
Using our theory, we can write an explicit form of the DM for these CSLs in terms of spin operators, which might provide new insight into their entanglement structure and {the topological features beyond the traditional single-particle description}.

\section*{Acknowledgments}
We are thankful to Weslei Fontana and Qian-Rui Lee for enlightening conversations during the completion of this work. 
This work is supported by the National Science and Technology Council, Taiwan under Grant No. 114-2112-M-001-026.
C.-C. W and Y.-P. H are supported under the MOST Young Scholar Fellowship (Grants No. 112-2636-M-007-008- No. 113-2636-M-007-002- and No. 114-2636-M-007-001-), National Center for Theoretical Sciences (Grants No. 113-2124-M-002-003-) from the Ministry of Science and Technology (MOST), Taiwan, and the Yushan Young Scholar Program (as Administrative Support Grant Fellow) from the Ministry of Education, Taiwan.

\appendix

\section{Projective Representation of Configuration Algebra}\label{appendix:projective-rep}

In this appendix, we rigorously discuss fundamental properties of the phase factor $\varphi(\mathfrak{D},\mathfrak{D}')$, defined in Eq.~\eqref{equ.Sigma-representation}, that emerges from the projective representation generated by string operators introduced in Sec.~\ref{sec:density-matrix-whole}. 

The discussions of this appendix are organized as follows. By Def.~\ref{def:Teven-odd}, we divide string operators into two groups according to their endpoint sets. Later, by Lemma~\ref{lemma:sigma-eta-relation}, we prove a fundamental property of string operators completely determined by the endpoint sets, elaborating the physical meaning of Def.~\ref{def:Teven-odd}. Theorems following Eq.~\eqref{eq:phi-phase-factor-def} discuss the algebra among string operators, and we conclude the appendix with Theorem~\ref{theorem:phi_exchange} and its corollary by proving that the algebra is equivalent to that of Majorana fermions, since both depend only on the endpoint sets of edge configurations. This also reinforces the conclusion made by Sec.~\ref{sec:Emergent-Fermions-at-the-Endpoints}. Notably, a similar mapping between string operators and Majorana fermions was also discussed in a recent study on quantum magic in the Kitaev model~\cite{Lamma-Bauer-Dalmonte-Collura_2026}.

\begin{definition}\label{def:Teven-odd}
  Given a class $[\mathfrak{D}]$, we denoted by $\mrm{EP}_a(\mathfrak{D})$ and $\mrm{EP}_b(\mathfrak{D})$  the set of endpoints in $A$-sublattice and $B$-sublattice, respectively. The difference in number between sublattice $A$ and $B$ is indicated by $\Delta(\mathfrak{D}) \equiv \abs{\mrm{EP}_a(\mathfrak{D})}-\abs{\mrm{EP}_b(\mathfrak{D})}$. A class is said to be in the set $\TReven$ if and only if $\abs{\mrm{EP}_a(\mathfrak{D})}-\abs{\mrm{EP}_b(\mathfrak{D})} \in 4\mathbb{Z}$. That is,
  \[
  \TReven := \lrg{[\mathfrak{D}] \; :\; \abs{\mrm{EP}_a(\mathfrak{D})}-\abs{\mrm{EP}_b(\mathfrak{D})} \in 4\mathbb{Z}} 
  \]
  Otherwise, it is in the set $\TRodd$.
  
  We further define a phase factor $\eta_{\mathfrak{D}} \equiv (-1)^{\Delta(\mathfrak{D})/2}$ to distinguish $\TReven$ and $\TRodd$, that is
  \begin{equation}\label{eq:eta-def}
      \eta_\mathfrak{D}\equiv (-1)^{\Delta(\mathfrak{D})/2} = \begin{cases}
                       +1, & [\mathfrak{D}]\in\TReven \\
                       -1, & [\mathfrak{D}]\in\TRodd
                     \end{cases}
  \end{equation}
\end{definition}

\begin{definition}\label{def:good-def}
  A class $[\mathfrak{D}]\in\TReven$ is said to be in a set $\Good$ if and only if $\abs{\mrm{EP}_a(\mathfrak{D})}=\abs{\mrm{EP}_b(\mathfrak{D})}$. That is,
  \[
  \Good = \lrg{[\mathfrak{D}] \; : \; \abs{\mrm{EP}_a(\mathfrak{D})} = \abs{\mrm{EP}_b(\mathfrak{D})}}
  \]
  $\mrm{EP}_a(\mathfrak{D})$ and $\mrm{EP}_b(\mathfrak{D})$ denote the set of endpoints in $A$-sublattice and $B$-sublattice, respectively. Note that the coefficients $\mathfrak{C}_{\mathfrak{D}}\neq 0$ iff $[\mathfrak{D}]\in\Good\subset\TReven$ due to Wick's theorem and the fact that $\anglelr{c_ic_j} = 0$ if both $i,j$ are in the same sublattice.
\end{definition}

In the following, we discuss the phase factor $\varphi(\mathfrak{D},\mathfrak{D}')$ of the projective representation of the configuration algebra and its relationship with the sign factor $\eta_\mathfrak{D}$ and the string operators $\Sigma_{\mathfrak{D}}$. First, we introduce a lemma that directly relates the string operators and the $\eta_{\mathfrak{D}}$ factors. We show that the sign factor $\eta_\mathfrak{D}$ is closely related to the equivalent Majorana representation of the edge configurations. 
\begin{lemma}\label{lemma:sigma-eta-relation}
The factor $\eta_\mathfrak{D}$ is determined by the square of the corresponding string operator as follows.
  \[
  \Sigma_\mathfrak{D}^2 = \eta_\mathfrak{D}\mathds{1}
  \]
\end{lemma}
\begin{figure}
    \centering
    \includegraphics[width=0.8\linewidth]{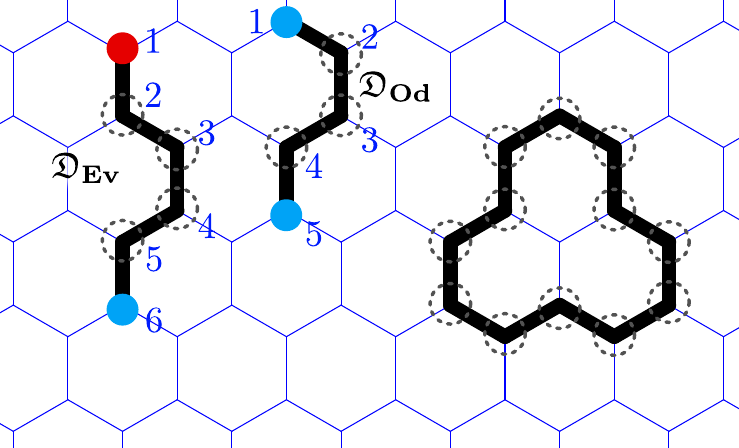}
    \caption{The number of joints determines the factor $\eta_\mathfrak{D}$ of a edge configuration $\mathfrak{D}$. Each joint, marked by dashed circles in the figure, contributes a factor $\pm i$ due to the algebra of Pauli spin operators. The string $\mathfrak{D}_{\bo{Ev}}$ has two endpoints in different sublattices, and there are an even number  (four, in this example) of joints on the string. In contrast, both endpoints of the string $\mathfrak{D}_{\bo{Od}}$ are in the same sublattice, and there are an odd number (three, in this example) of joints on the string. A loop, located in the right portion of the figure, doesn't have endpoints, and there are always an even number (twelve in this example) of joints on a closed loop.
    }
    \label{fig:joints}
\end{figure}
\begin{proof}
    Let $\mathsf{S}$ be an element of the projective Pauli group $\mathcal{P}_N \equiv \lrg{\mathds{1}, \sigma^x, \sigma^y, \sigma^z}^{\otimes N}$ of size $N$, which is the traditional Pauli group without phases $\lrg{\pm 1, \pm i}$. We define $\mathsf{S}_{\mathfrak{D}} \in \mathcal{P}_N$ as an operator that is proportional to the string operator of the edge configuration $\mathfrak{D}$, i.e., $\mathsf{S}_\mathfrak{D} = \omega_\mathfrak{D} \Sigma_\mathfrak{D}$ with a phase $\omega_\mathfrak{D}\in\lrg{\pm1, \pm i}$ arising from the algebra of Pauli spin operators. According to the definition of the string operator given by Eq.~\eqref{equ.string-operator-def}
    and the structure of strings illustrated in \autoref{fig:joints}, each joint on the string contributes a $\pm i$ to the phase factor $\omega_\mathfrak{D}$ due to the algebra $\sigma^\mu\sigma^\nu = i\epsilon^{\mu\nu\lambda}\sigma^\lambda$. Owing to this, the following relation holds for $\omega_\mathfrak{D}$,
    \[
    \omega_\mathfrak{D}^2 = \lrm{(\pm i)^{2}}^{\text{No. of joints on $\mathfrak{D}$}} = (-1)^{\text{No. of joints on $\mathfrak{D}$}}.
    \]
    Combining this result and the fact that every element in $\mathcal{P}_N$ squares to one
    \[
    \mathsf{S}_{\mathfrak{D}}^2 = \omega_\mathfrak{D}^2\Sigma_\mathfrak{D}^2  = \mathds{1},
    \]
    we conclude that 
    \begin{equation}\label{eq:sigma-joints}
        \Sigma_\mathfrak{D}^2 = (-1)^{\text{No. of joints on $\mathfrak{D}$}} \mathds{1}.
    \end{equation}
    
    Next, we show that the number of joints determines $\eta_\mathfrak{D}$, which leads to the theorem. Edge configurations $\mathfrak{D}$ consisting of a single connected string with two endpoints can be divided into two types. The first type, which is a subset of $\TReven$, consists of strings with two endpoints in different sublattices, while the second type, which is a subset of $\TRodd$, consists of strings with both endpoints in the same sublattice. According to \autoref{fig:joints}, the strings in $\TReven$ have an even number of joints; hence, the corresponding phase factor is $\eta_\mathfrak{D} = 1$. In contrast, the strings in $\TRodd$ have an odd number of joints; hence, the corresponding phase factor is $\eta_\mathfrak{D} = -1$. Consider a $\TReven$-edge configuration with $\abs{\mrm{EP}_a(\mathfrak{D})} - \abs{\mrm{EP}_b(\mathfrak{D})} \in 4\mathbb{Z}$, an integer-multiple-of-4 endpoints of the same sublattice type come from an integer-multiple-of-2 strings possessing two endpoints in the same sublattice. Therefore, the negative $\eta$-factors from these strings cancel each other, leaving the overall phase $\eta_\mathfrak{D} = +1$. In contrast, if $\mathfrak{D} \in \TRodd$, the negative phases cannot be canceled and the overall phase will be left with a negative sign $\eta_\mathfrak{D} = -1$. Together, the above analysis and Definition~\eqref{eq:eta-def} imply that $\eta_\mathfrak{D}$ is determined by the number of joints on the string,
    \begin{equation}\label{eq:eta-joints}
        \eta_\mathfrak{D} = (-1)^{\text{No. of joints on $\mathfrak{D}$}} 
    \end{equation}
    Substituting \eqref{eq:eta-joints} into \eqref{eq:sigma-joints} completes the proof.
\end{proof}

\begin{corollary}\label{corollary:good-square}
    A string operator $\Sigma_\mathfrak{D}$ squares to one if $\mathfrak{C}_\mathfrak{D} \neq 0$ in the pure Kitaev ground state.
\end{corollary}
\begin{proof}
    As mentioned in Definition~\ref{def:good-def}, a necessary condition for an edge configuration $\mathfrak{D}$ to have a finite correlation function $\mathfrak{C}_\mathfrak{D} \neq 0$ is $[\mathfrak{D}] \in \Good$. Because $\Good\subset\TReven$, the phase factor of $[\mathfrak{D}] \in \Good$ must be positive, $\eta_\mathfrak{D} = +1$. Due to this and Lemma~\ref{lemma:sigma-eta-relation}, we have 
    \[
    \Sigma_\mathfrak{D}^2 = \eta_\mathfrak{D}\mathds{1} = \mathds{1} \qquad \text{if $[\mathfrak{D}] \in \Good$}.
    \]
    This completes the proof.
\end{proof}

Lemma~\ref{lemma:sigma-eta-relation} with Eq.~\eqref{eq:eta-joints} implies that the string operators are equivalent to the Majorana fermion representation as illustrated by Eqs.~\eqref{eq.S-rep-swap} and \eqref{eq.M-rep_swap}. Here we give a simple example to illustrate the idea. The string $\mathfrak{D}_{\bo{Ev}}$ illustrated in Fig.~\ref{fig:joints} represented in the Majorana fermionic representation given by Eq.~\eqref{equ.V-rep-def} reads
\begin{align*}
\mathcal{M}_{\mathfrak{D}_{\bo{Ev}}}
&=
\lr{ic_2c_1}\lr{ic_2c_3}\lr{ic_4c_3}\lr{ic_4c_5}\lr{ic_6c_5} \\
&= -i^{1 + \text{No. of joints}}c_1c_6 \\
&= i^{\text{No. of joints}} \lr{-ic_1c_6}.
\end{align*}
This derivation shows that Majorana fermions at internal joints cancel out in pairs, as adjacent dimers share the same site. The resulting operator depends only on the fermions at the endpoints and a phase determined by the number of joints of the string. Since $\lr{-ic_1c_6}^2 = 1$, we obtain
\[
\mathcal{M}_{\mathfrak{D}}^2 = (-1)^{\text{No. of joints on $\mathfrak{D}$}} = \eta_{\mathfrak{D}},
\]
where the result has been generalized to arbitrary strings. This form resembles Lemma~\ref{lemma:sigma-eta-relation}, and we will prove that the algebras of the spin and Majorana fermion representations are identical by discussing the algebra of the string operators. 

The string operators form a projective representation of the algebra $\anglelr{\lrg{\mathfrak{D}}, \ominus}$ for edge configurations as follows,
\begin{equation}\label{eq:phi-phase-factor-def}
  \Sigma_{\mathfrak{D}}\Sigma_{\mathfrak{D}'} = \varphi(\mathfrak{D},\mathfrak{D}') \Sigma_{\mathfrak{D}\ominus\mathfrak{D}'},
\end{equation}
where $\varphi(\mathfrak{D},\mathfrak{D}')\in\lrg{-1,1}$ is the phase factor of the projective representation. 
\begin{theorem}\label{cor:phase-associativity}
  Due to the associativity of the representation, the following relation holds for three arbitrary edge configurations $\mathfrak{D}$, $\mathfrak{D}'$, and $\mathfrak{D}''$,
\[
\varphi(\mathfrak{D},\mathfrak{D}'\ominus\mathfrak{D}'')\varphi(\mathfrak{D}',\mathfrak{D}'') = 
\varphi(\mathfrak{D},\mathfrak{D}')\varphi(\mathfrak{D}\ominus\mathfrak{D}',\mathfrak{D}'').
\]
\end{theorem}

\begin{theorem}\label{thm:phi-operational-def}
The phase factor defined in Eq.~\eqref{eq:phi-phase-factor-def} can be reformulated as follows,
\[
\varphi(\mathfrak{D},\mathfrak{D}') = \frac{1}{2^{N}}\eta_{\mathfrak{D}\ominus\mathfrak{D}'}\Tr\lrm{\Sigma_{\mathfrak{D}}\Sigma_{\mathfrak{D}'}\Sigma_{\mathfrak{D}\ominus\mathfrak{D}'}}.
\]
\end{theorem}
\begin{proof}
First, convert the multiplication of the first two string operators $\Sigma_\mathfrak{D}$ and $\Sigma_{\mathfrak{D}'}$ into the string operator $\Sigma_{\mathfrak{D}\ominus\mathfrak{D}'}$ of the resulting configuration using the definition of the phase factor Eq.~\eqref{eq:phi-phase-factor-def}
\[
\Tr\lrm{\Sigma_{\mathfrak{D}}\Sigma_{\mathfrak{D}'}\Sigma_{\mathfrak{D}\ominus\mathfrak{D}'}}
= \varphi(\mathfrak{D},\mathfrak{D}')\Tr\lrm{\Sigma_{\mathfrak{D}\ominus\mathfrak{D}'}\Sigma_{\mathfrak{D}\ominus\mathfrak{D}'}}.
\]
According to Lemma~\ref{lemma:sigma-eta-relation}, the square of the string operator in the trace reduces to the $\eta$ factor
\[
\Tr\lrm{\Sigma_{\mathfrak{D}\ominus\mathfrak{D}'}\Sigma_{\mathfrak{D}\ominus\mathfrak{D}'}}
=
\Tr\lr{\mathds{1}}\eta_{\mathfrak{D}\ominus\mathfrak{D}'}.
\]
Combining the two equations completes the proof.
\end{proof}

\begin{lemma}\label{thm:phi-self}
 Given an arbitrary edge configuration $\mathfrak{D}$, the following two relations hold.
 
\begin{enumerate}
    \item [(i)] $\varphi(\varnothing, \mathfrak{D}) = \varphi(\mathfrak{D},\varnothing)= 1$
    \item [(ii)] $\varphi(\mathfrak{D},\mathfrak{D}) = \eta_{\mathfrak{D}}$
\end{enumerate}
\begin{proof}
    The first relation comes from Eq.~\eqref{eq:phi-phase-factor-def} and the fact that $\Sigma_\varnothing = \mathds{1}$. The second relation is equivalent to Lemma~\ref{lemma:sigma-eta-relation} reformulated using the factor $\varphi$ given in Eq.~\eqref{eq:phi-phase-factor-def}.
\end{proof}
\end{lemma}

\begin{lemma}\label{g-reduce-theorem.thm}
Given two arbitrary edge configurations $\mathfrak{D}$ and $\mathfrak{D}'$, the following two equations hold.
\begin{enumerate}
    \item [(i)] $\varphi(\mathfrak{D},\mathfrak{D}\ominus\mathfrak{D}') = \eta_{\mathfrak{D}'}\eta_{\mathfrak{D}\ominus\mathfrak{D}'} \varphi(\mathfrak{D}',\mathfrak{D})$
    \item [(ii)] $\varphi(\mathfrak{D}\ominus\mathfrak{D}',\mathfrak{D}) = \eta_{\mathfrak{D}'}\eta_{\mathfrak{D}\ominus\mathfrak{D}'} \varphi(\mathfrak{D},\mathfrak{D}')$.
\end{enumerate}
\end{lemma}
\begin{proof}
Applying Theorem~\ref{thm:phi-operational-def} directly, we have
  \begin{align*}
    \varphi(\mathfrak{D},\mathfrak{D}\ominus\mathfrak{D}') &= \frac{\eta_{\mathfrak{D}\ominus\mathfrak{D}\ominus\mathfrak{D}'}}{2^{N}}\Tr\lrm{\Sigma_{\mathfrak{D}}\Sigma_{\mathfrak{D}\ominus\mathfrak{D}'}\Sigma_{\mathfrak{D}\ominus\mathfrak{D}\ominus\mathfrak{D}'}} \\
    &= \frac{\eta_{\mathfrak{D}'}}{2^{N}}\Tr\lrm{\Sigma_{\mathfrak{D}}\Sigma_{\mathfrak{D}\ominus\mathfrak{D}'}\Sigma_{\mathfrak{D}'}} \\
    &=\eta_{\mathfrak{D}'}\eta_{\mathfrak{D}'\ominus\mathfrak{D}} \varphi(\mathfrak{D}',\mathfrak{D}).
  \end{align*}
The result proves the first equation of the lemma. A similar approach leads to the second equation of the lemma.
\end{proof}
\begin{lemma}\label{g_reduce_symmetric.e}
Given two arbitrary edge configurations $\mathfrak{D}$ and $\mathfrak{D}'$, the following relation holds,
  \[
  \varphi(\mathfrak{D}, \mathfrak{D}\ominus\mathfrak{D}') = \eta_{\mathfrak{D}}\varphi(\mathfrak{D},\mathfrak{D}').
  \]
\end{lemma}
\begin{proof}
  According to Theorem~\ref{cor:phase-associativity}, the associativity of the projective representation implies
  \[
  \varphi(\mathfrak{D},\mathfrak{D}\ominus\mathfrak{D}')\varphi(\mathfrak{D},\mathfrak{D}') = \varphi(\mathfrak{D},\mathfrak{D})\varphi(\mathfrak{D}\ominus\mathfrak{D},\mathfrak{D}').
  \]
  According to Lemma \ref{thm:phi-self}, the terms on the right-hand side of the equation can be replaced by $\eta$ factors. Therefore, the above relation reduces to the following,
  \[
  \varphi(\mathfrak{D},\mathfrak{D}\ominus\mathfrak{D}')\varphi(\mathfrak{D},\mathfrak{D}') = \eta_{\mathfrak{D}}.
  \]
  Since $\varphi,\eta = \pm 1$, we obtain
  \[
  \varphi(\mathfrak{D},\mathfrak{D}\ominus\mathfrak{D}') = \eta_{\mathfrak{D}}\varphi(\mathfrak{D},\mathfrak{D}').
  \]
  This completes the proof.
\end{proof}
\begin{theorem}\label{theorem:phi_exchange}
Given two arbitrary edge configurations $\mathfrak{D}$ and $\mathfrak{D}'$, the sign factor produced by exchanging the positions of the two configurations in $\varphi(\mathfrak{D},\mathfrak{D}')$ is determined by the configuration of endpoints
  \begin{equation}\label{eq:exchange-relation-phi}
      \varphi(\mathfrak{D},\mathfrak{D}')
  =
  (-1)^{\abs{\mrm{EP}(\mathfrak{D})\cap \mrm{EP}(\mathfrak{D}')}} \varphi(\mathfrak{D}',\mathfrak{D}),
  \end{equation}
  where $\mrm{EP}(\mathfrak{D})$ denotes the set of endpoints of the edge configuration $\mathfrak{D}$.
\end{theorem}
\begin{proof}
  Combining Lemmas \ref{g-reduce-theorem.thm} and \ref{g_reduce_symmetric.e} leads to the following equation,
  \[
  \varphi(\mathfrak{D},\mathfrak{D}\ominus\mathfrak{D}') = 
  \eta_{\mathfrak{D}'}\eta_{\mathfrak{D}\ominus\mathfrak{D}'} \varphi(\mathfrak{D}',\mathfrak{D}) =
  \eta_{\mathfrak{D}}\varphi(\mathfrak{D},\mathfrak{D}').
  \]
  Moving the factor $\eta_\mathfrak{D}$ in the last term to the left, we obtain
  \begin{equation}\label{eq:eta-phi-relation}
      \varphi(\mathfrak{D},\mathfrak{D}') = \eta_{\mathfrak{D}}\eta_{\mathfrak{D}'}\eta_{\mathfrak{D}\ominus\mathfrak{D}'} \varphi(\mathfrak{D}',\mathfrak{D}) .
  \end{equation}
  According to Definition~\ref{eq:eta-def}, the phase factor $\eta_{\mathfrak{D}}\eta_{\mathfrak{D}'}\eta_{\mathfrak{D}\ominus\mathfrak{D}'}$ can be represented in terms of the number of endpoints.
  \begin{equation}\label{eq:eta-phi-proof-1}
      \eta_{\mathfrak{D}}\eta_{\mathfrak{D}'}\eta_{\mathfrak{D}\ominus\mathfrak{D}'}
  =
  (-1)^{[\Delta(\mathfrak{D}) + \Delta(\mathfrak{D}') + \Delta(\mathfrak{D}\ominus\mathfrak{D}')]/2}.
  \end{equation}
  The term $\Delta(\mathfrak{D}\ominus\mathfrak{D}')$ in the exponent of the above equation can be decomposed into the following form,
  \begin{align*}
      \Delta(\mathfrak{D}\ominus\mathfrak{D}') 
      &=
      \abs{\mrm{EP}_a(\mathfrak{D}\ominus\mathfrak{D}')} - \abs{\mrm{EP}_b(\mathfrak{D}\ominus\mathfrak{D}')} \\
      &= 
      \Delta(\mathfrak{D}) + \Delta(\mathfrak{D}') - 2\abs{\mrm{EP}_a(\mathfrak{D}) \cap \mrm{EP}_a(\mathfrak{D}')} \\
          & \hspace{5em} + 2\abs{\mrm{EP}_b(\mathfrak{D}) \cap \mrm{EP}_b(\mathfrak{D}')} \\
      &=
      \Delta(\mathfrak{D}) + \Delta(\mathfrak{D}') - 2\abs{\mrm{EP}(\mathfrak{D}) \cap \mrm{EP}(\mathfrak{D}')} \\
          & \hspace{5em} + 4\abs{\mrm{EP}_b(\mathfrak{D}) \cap \mrm{EP}_b(\mathfrak{D}')}.
  \end{align*}
  Substitute this form into Eq.~\eqref{eq:eta-phi-proof-1}, we obtain,
  \begin{align*}
      \eta_{\mathfrak{D}}\eta_{\mathfrak{D}'}\eta_{\mathfrak{D}\ominus\mathfrak{D}'}
      &=
      (-1)^{
        \lrm{
            \scriptsize
        \begin{aligned}
            &\Delta(\mathfrak{D}) + \Delta(\mathfrak{D}') - 
            \abs{\mrm{EP}(\mathfrak{D}) \cap \mrm{EP}(\mathfrak{D}')} \\
            & \qquad  + 2 \abs{\mrm{EP}_b(\mathfrak{D}) \cap \mrm{EP}_b(\mathfrak{D}')}
        \end{aligned}
        }
      } \\
      &= (-1)^{\abs{\mrm{EP}(\mathfrak{D}) \cap \mrm{EP}(\mathfrak{D}')}}.
  \end{align*}
  With this result, Eq.~\eqref{eq:eta-phi-relation} reduces to the form given in the theorem.
\end{proof}
\begin{corollary}
The phase factor arising from the exchange of two string operators is determined by the configuration of the string endpoints as follows.
  \begin{equation}\label{eq:exchange-relation}
      \Sigma_{\mathfrak{D}}\Sigma_{\mathfrak{D}'} = (-1)^{\abs{\mrm{EP}(\mathfrak{D}) \cap \mrm{EP}(\mathfrak{D}')}} 
  \Sigma_{\mathfrak{D}'}\Sigma_{\mathfrak{D}}
  \end{equation}
\end{corollary}
In the fermionic representation, the above equation represents the anti-commutation among fermions that emerge at the endpoints of strings. Eq.~\eqref{eq:exchange-relation} is equivalent to the following well-known relation in Majorana fermion codes~\cite{Bravyi_2010_majorana-code}
\[
c_A c_B = (-1)^{\abs{A}\cdot\abs{B} + \abs{A\cap B}}c_Bc_A,
\]
where $c_R \equiv \prod_{i\in R}c_i$ is the assembly of Majorana fermions in the region $R$. The above relation for Majorana fermions reduces to Eq.~\eqref{eq:exchange-relation} if either $\abs{A}$ or $\abs{B}$ is even, which automatically holds for strings since they always have an even number of endpoints.

\section{Fermion Parity from String Operators}\label{sec:fermion_parity_from_string_operators}

The physical meaning of the total fermion parity in the spin representation is important, not only because the subtlety about the definition of equivalence classes arises from it, but also because of its importance when calculating the reduced density matrix.

\begin{figure}
    \centering
    \includegraphics[width=0.48\linewidth]{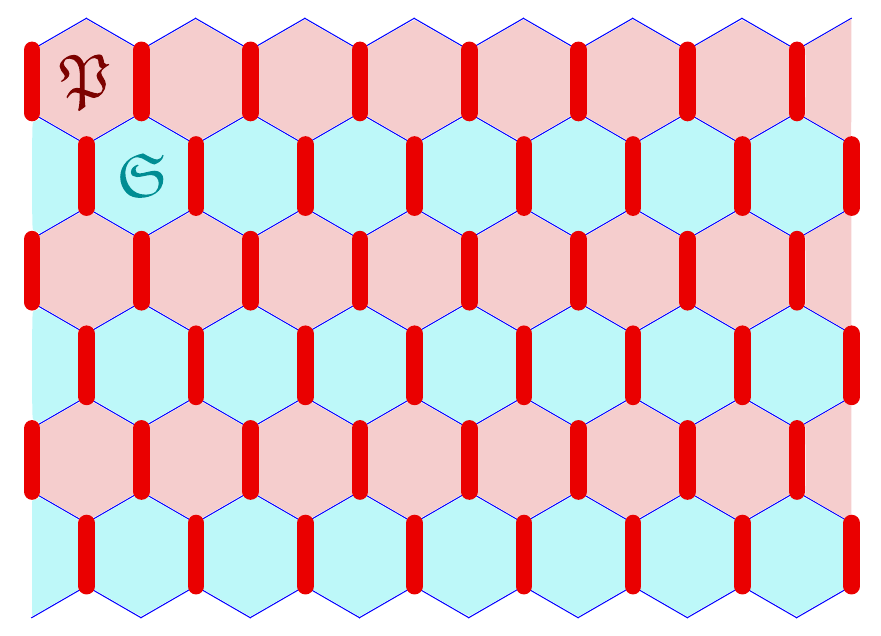}
    \includegraphics[width=0.48\linewidth]{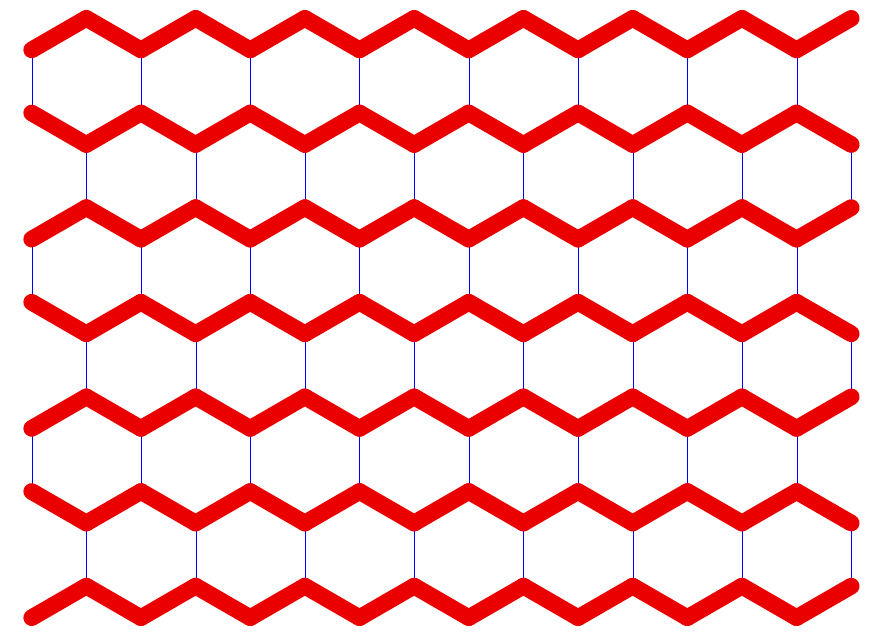}
    \caption{The left panel illustrates the plaqeutte configurations $\mathfrak{P}$ and $\mathfrak{S}$ as well as the edge configuration $\mathfrak{D}_Z$. The set of plaquettes in pink is denoted by $\mathfrak{P}$, while the set containing those in pink-blue is denoted by $\mathfrak{S}$. The edge configuration $\mathfrak{D}_Z$ consists of all of the red $z$-directional edges in the system, as shown in the above figure. The right panel illustrates the zigzag configuration $\mathfrak{D}_{zz}$ consisting of edges at all the horizontal loops. Such an edge configuration is also the boundary of either $\mathfrak{P}$ or $\mathfrak{S}$, for which we can write $\mathfrak{D}_{zz} = \de\mathfrak{P} = \de\mathfrak{S}$. Note that $\Sigma_{\mathfrak{D}_Z}=\Sigma_{\mathfrak{D}_{zz}}$ although $\mrm{EP}(\mathfrak{D}_Z)\neq \mrm{EP}(\mathfrak{D}_{zz})$} 
    \label{fig.S-P-dimer-gas}
\end{figure}

String operators of edge configurations in the same class have the same magnitude of expectation values,
\[
\abs{\anglelr{\Sigma_{\mathfrak{D}}}} = \abs{\anglelr{\Sigma_{\mathfrak{D}'}}} \quad\text{if $\mathfrak{D}\sim\mathfrak{D}'$}
\]
hence contribute the same weight to the density matrix. Nevertheless, it is still possible for two configurations from different classes to have the same weight contribution to the density matrix. There are a type of plaquette configurations, $\mathfrak{P}$ and $\mathfrak{S}$, that transform an edge configuration $\mathfrak{D}$ to $\mathfrak{D}\ominus\mathfrak{D}_Z$ because of the following relation
\begin{equation}\label{equ.P-S-dimer}
    \prod_{p\in\mathfrak{P}}W_p = \prod_{p\in\mathfrak{S}}W_p = \prod_{\anglelr{ij}\in \mathfrak{D}_Z}\sigma_i^z\sigma_{j}^z = \Sigma_{\mathfrak{D}_Z}.
\end{equation}
The plaquette configurations $\mathfrak{P}$ and $\mathfrak{S}$ as well as the edge configuration $\mathfrak{D}_Z$ are illustrated in the left panel of \autoref{fig.S-P-dimer-gas}. The edge configuration $\mathfrak{D}_Z$ is composed of the $z$-directional dimers all over the system. The plaquette configuration $\mathfrak{P}$ contains all the plaquettes at the even horizontal layers, while $\mathfrak{S}$ is the complement of $\mathfrak{P}$. The endpoint configuration $\mrm{EP}(\mathfrak{D}_Z)$ is simply the set of all the lattice sites in the system. On the other hand, the operators $\prod_{p\in\mathfrak{P}}W_p$ and $\prod_{p\in \mathfrak{S}}W_p$ is also equal to the string operator of the zigzag edge configuration $\mathfrak{D}_{zz}$, a set of all the $x$- and $y$-edges illustrated in the right panel of \autoref{fig.S-P-dimer-gas}, because $\de\mathfrak{P} = \de\mathfrak{S} = \mathfrak{D}_{zz}$
\begin{equation}\label{equ.M-MZ-complement}
\prod_{p\in\mathfrak{P}}W_p = \prod_{p\in\mathfrak{S}}W_p  = \Sigma_{\mathfrak{D}_{zz}}.
\end{equation}
Unlike $\mathfrak{D}_Z$, this edge configuration has no endpoints, $\mrm{EP}(\mathfrak{D}_{zz})=\varnothing$. So, we must keep in mind that any two configurations, $\mathfrak{D}\ominus\mathfrak{D}_Z$ and $\mathfrak{D}\ominus\mathfrak{D}_{zz}$, give an identical string operator, $\Sigma_{\mathfrak{D}\ominus\mathfrak{D}_Z} = \Sigma_{\mathfrak{D}\ominus\mathfrak{D}_{zz}}$, although they are not equivalent.

We would also like to show that the physical meaning of the operator $\Sigma_{\mathfrak{D}_Z}$ is the total fermion parity of the $c$-Majorana fermions. Because the expectation values of fluxes are fixed to be $\anglelr{W_p}=+1$ in the ground state, we directly know that the expectation value of $\Sigma_{\mathfrak{D}_Z}$ is equal to $+1$ directly from Eq.~\eqref{equ.P-S-dimer} 
\begin{equation}\label{equ.fermion-parity-dimer-rep}
    \anglelr{\Sigma_{\mathfrak{D}_Z}}=1
\end{equation}

We can understand Eq.~\eqref{equ.fermion-parity-dimer-rep} through fermion parity. In our convention, the gauge is fixed as $u_{ij}=1\forall\anglelr{ij}$, therefore, the string operator $\Sigma_{\mathfrak{D}_Z}$ can be transformed into the fermion parity operator $\hat{\bog{\pi}}$~\footnote{Two distinct Majorana fermions, say $c_1$ and $c_2$, can combine to form a normal fermion degree of freedom $f$ in a way that $f=\frac{1}{2}\lr{c_1+ic_2}$ and $f^\dagger=\frac{1}{2}\lr{c_1-ic_2}$. In this way, the fermion parity operator represented using the Majorana fermions becomes $\hat{\bog{\pi}}\equiv(-1)^{f^\dagger f}=ic_1c_2$. In a more general sense, the fermion parity operator is equal to the multiplication of all the Majorana fermions $\prod_i c_i$ up to a phase.}
\begin{equation}\label{equ.total-fermion-parity-Sigma-Z}
    \Sigma_{\mathfrak{D}_Z} \to \prod_{\anglelr{ij}\in\mathfrak{D}_Z}ic_ic_j= i^{\abs{\Lambda}/2} \prod_{i\in\Lambda}c_i = \hat{\bog{\pi}}.
\end{equation} 
From the Majorana fermion point of view, the fermion representation of the Kitaev Hamiltonian \eqref{equ.MF-Hamiltonian} is quadratic. As a consequence, the total fermion parity of the whole system should be conserved and must be $+1$ on a torus~\cite{Fu-Knolle-Perkins-2018}. This fermion-parity degree of freedom is emergent only in the Majorana representation and should be fixed by the intrinsic property of the spin operators in the physical spin representation. This is why the expectation value of the operator $\Sigma_{\mathfrak{D}_Z}$ is $+1$. 

\section{Explicit Calculation of Reduced Density Matrices}\label{appendix:reduced-density0matrix}
We follow the general reduced density matrix of a simply-connected region $A$ and continue to carry out the partial trace explicitly to derive the reduced density matrix. Instead of isolating $\prod_{p\in A}\lr{\mathds{1}+W_p}$ at the beginning, it is clearer to see what terms survive after tracing out the environment if we include the whole part of the density matrix while calculating the trace. Both $\hat{\Pi}$ and $\hat{\Pi}_A$ are projectors, and $\hat{\Pi}$ is a projector onto a subspace of the Hilbert space to which $\hat{\Pi}_A$ projects. From this observation, we can duplicate the projector $\hat{\Pi}_A$ as a redundant term: $\hat{\Pi} = \hat{\Pi}\hat{\Pi}_A$. Therefore, the reduced density matrix of a simply-connected region $A$ can be written as
\begin{align*}
\rho_A &= \frac{1}{2^{N-N_p+2}}\Tr_{\overline{A}}\lrm{\hat{\Pi}_A\hat{\Pi}\sum_{[\mathfrak{D}]}\mathfrak{C}_\mathfrak{D}\Sigma_\mathfrak{D}} \\
&= 
\frac{1}{2^{N+2}}\hat{\Pi}_A \sum_{[\mathfrak{D}]}\mathfrak{C}_\mathfrak{D} \Tr_{\overline{A}}\lrm{\prod_{p}\lr{\mathds{1}+W_p}\Sigma_\mathfrak{D}}
\end{align*}
After that, we investigate what terms can survive after the following trace,
\begin{align*}
    &\sum_{[\mathfrak{D}]}\mathfrak{C}_{\mathfrak{D}}\Tr_{\overline{A}}\lrm{\prod_{p}\lr{\mathds{1} + W_p}\Sigma_{\mathfrak{D}}}\\
=&
2\sum'_{\mathcal{A}}\sum_{[\mathfrak{D}]}\mathfrak{C}_{\mathfrak{D}} \Tr_{\overline{A}}\lrm{ \lr{\prod_{p\in\mathcal{A}}W_p }\Sigma_{\mathfrak{D}}}
\end{align*}
As a reminder, we denote by $\mathcal{A}$ the sets of plaqettes as defined in Eq.~\eqref{equ.lattice-Stokes}. Since only two types of terms survive after taking the trace, which are those with (i) $\mathcal{A}\subset A$ and $[\mathfrak{D}]\subset A$ and (ii) $\mathcal{A}\ominus\mathfrak{P}\subset A$ and $[\mathfrak{D}\ominus\mathfrak{D}_Z]\subset A$, respectively. Both give the same contribution. The reduced density matrix is therefore
\begin{align*}
&2\sum'_{\mathcal{A}}\sum_{[\mathfrak{D}]}\mathfrak{C}_{\mathfrak{D}} \Tr_{\overline{A}}\lrm{ \lr{\prod_{p\in\mathcal{A}}W_p } \Sigma_{\mathfrak{D}}} \\
=& 2\prod_{p\in A}\lr{\mathds{1}+W_p}\lrm{ 2\sum_{[\mathfrak{D}]\subset A} \mathfrak{C}_{\mathfrak{D}}\Sigma_{\mathfrak{D}} 
}
\Tr_{\overline{A}}\lr{\mathds{1}_{\overline{A}}} \\
=& \hat{\Pi}_A 2^{N-N^A+N^A_p+2} \sum_{[\mathfrak{D}]\subset A} \mathfrak{C}_{\mathfrak{D}}\Sigma_{\mathfrak{D}}
\end{align*}
Substituting this result back into the density matrix, we finally have
\begin{equation}\label{GeneralReducedDensityMatrix_first.e}
  \rho_A = \frac{1}{2^{N^A-N^A_p}}\hat{\Pi}_A \sum_{[\mathfrak{D}]\subset A}\mathfrak{C}_\mathfrak{D} \Sigma_{\mathfrak{D}}
\end{equation}
The result is in agreement with Eq.~\eqref{equ.red-den-mat-sim-con}, which has been verified to be properly normalized and to yield correct expectation values for observables.

\section{Fermion Representation of the Density Matrix}\label{appendix:Majorana-derivation}

The ground state of Hamiltonian \eqref{equ.MF-Hamiltonian} reads~\cite{Yao-Qi_2010}
\begin{equation}\label{equ.fermion-state}
    \ket{\psi} = \sqrt{2^{N-1}} \mathcal{P}_G\ket{u}\ket{M}
\end{equation}
The Gutzwiller projector $\mathcal{P}_G=\prod_{i}\lr{\frac{\mathds{1}+D_i}{2}}$ implements the Gauss law of the $\mathbb{Z}_2$ gauge theory in the Majorana representation of the Kitaev model, and $D_i = b^x_ib^y_ib^z_i c_i$ is the generator of the $\mathbb{Z}_2$ gauge transform. {Aside from the gauge theory perspective, the physical meaning of a Gutzwiller projection is to project states in the enlarged parton Hilbert space back to the physical space. As mentioned in Sec.~\ref{sec:model-intro}, the Pauli spin operators must satisfy $-i\sigma^x\sigma^y\sigma^z = \mathds{1}$. Substituting the Majorana representation of the Pauli spin operators into this relation leads to the physical constraint for partons $D_i = b^x_ib^y_ib^z_i c_i = \mathds{1}$ in the physical sector.} 

The Gutzwiller projector can also be expanded in the following way~\cite{Pedrocchi-Chesi-Loss_2011, Zschocke-Vojta_2015, Fu-Knolle-Perkins-2018}
\[
\mathcal{P}_G = \prod_{i}\lr{\frac{\mathds{1} + D_i}{2}}= \frac{1}{2^{N-1}}\lr{\frac{\mathds{1} + \prod_iD_i}{2}}\sum_{\lrg{i}}'\prod_{j\in\lrg{i}}D_j
\]
where the primed sum stands for a summation over subsets $\lrg{i}$ with sizes smaller than half of the system, $\abs{\lrg{i}}\leq N/2$. The coefficient $\sqrt{2^{N-1}}$ in the state \eqref{equ.fermion-state} is the factor that normalizes the state
\begin{align*}
    \sbraket{\psi}{\psi} &= 2^{N-1}\bra{M}\bra{u}\mathcal{P}_G\ket{u}\ket{M} \\
    &= \bra{M}\bra{u}\lr{\frac{\mathds{1} + \prod_iD_i}{2}}\sum_{\lrg{i}}'\prod_{j\in\lrg{i}}D_j\ket{u}\ket{M} \\
    &= \sum_{\lrg{i}}'\bra{M}\bra{u}\prod_{j\in\lrg{i}}D_j\ket{u}\ket{M} \\
    &= 1
\end{align*}
In the last second line, we used a fact that the state $\ket{u}\ket{M}$ is intrinsically an eigenvector of the operator $\prod_i D_i$ with eigenvalue $+1$. This operator is $\mathbb{Z}_2$ and commutes with the Hamiltonian; therefore, the state $\ket{u}\ket{M}$ is an eigenvector with eigenvalue $\pm 1$. The projector $\lr{\frac{\mathds{1}+\prod_iD_i}{2}}$, hidden in the Gutzwiller projector, selects only the states with eigenvalue $ +1$. Therefore, we only pick the state $\ket{u}\ket{M}$ with the eigenvalue $+1$. As for the last line of the derivation, we know that $D_i$ is the generator of the $\mathbb{Z}_2$ gauge transform and can flip the gauge fields residing on the three edges connected to the site $i$, which implies that the state $D_i\ket{u}$ is orthogonal to the original state $\ket{u}$ before being flipped, so only the term $\lrg{i}=\varnothing$ survives in the summation. It is worth mentioning that the operator is nothing but the fermion parity operator $\hat{\bog{\pi}}$ multiplied with all of the gauge field operators up to a phase~\cite{Pedrocchi-Chesi-Loss_2011}
\[
\prod_i D_i = \prod_ib^x_ib^y_ib^z_ic_i \propto \prod_{\anglelr{ij}}b^{\alpha_{ij}}_{i}b^{\alpha_{ij}}_j\prod_ic_i \propto \lr{\prod_{\anglelr{ij}}\hat{u}_{ij}}\hat{\bog{\pi}}
\]
The phase could be $\pm 1$, depending on the boundary condition of the system~\cite{Pedrocchi-Chesi-Loss_2011, Zschocke-Vojta_2015}. Since $\ket{u}$ is the eigenvector of gauge field operators $\hat{u}_{ij}$ and $\ket{M}$ is the eigenvector of $\hat{\bog{\pi}}$ as the total fermion parity is fixed, it is natural that the state $\ket{u}\ket{M}$ is the eigenvector of the operator $\prod_i D_i$ with the eigenvalue $\pm 1$.

Now, we construct the DM in the Majorana-fermion representation. The DM of the ground state is a pure state of $\ket{\psi}$
\[
\rho = \ket{\psi}\bra{\psi} = 2^{N-1}\mathcal{P}_G \lr{\ket{u}\bra{u}\otimes\ket{M}\bra{M}} \mathcal{P}_G
\]
The gauge part $\ket{u}\bra{u}$ is a projector onto the state $\ket{u}$ obtained by fixing the gauge. As usual, for a ground state, the gauge can be fixed trivially as $u_{ij}=+1\forall\anglelr{ij}$. Thus, the projector $\ket{u}\bra{u}$ in such a gauge can be expressed as
\begin{equation}
    \ket{u}\bra{u} = \prod_{\anglelr{ij}}\lr{\frac{\mathds{1} + \hat{u}_{ij}}{2}}
\end{equation}
The projector $\ket{M}\bra{M}$ is the DM of the ground state of the effective free Majorana fermion system, and is denoted by $\ket{M}\bra{M}=\rho^F$ in the main part of this paper. The DM $\rho^F$ only contains the information of the matter sector, and can be expressed in terms of those $c$-Majorana fermion operators. The form of $\rho_F$ can be derived in a way similar to Sec.~\ref{sec:density-matrix-calculate}. In a system of Majorana fermions, the elements that make up the operators are the Majorana fermion operators. Therefore, the DM can be expanded in the form
\[
\rho_F = \frac{1}{2^{N/2}}\sum_{\lrg{i}} \anglelr{\prod_{i\in\lrg{i}}c_i}^* \prod_{i\in\lrg{i}}c_i,
\]
where $\lrg{i}$ stands for the set of lattice points. The normalization factor $1/2^{N/2}$ comes from the Hilbert space dimension of a Majorana fermion system. Furthermore, a product of Majorana fermions $\prod_{i\in\lrg{i}}c_i$ is equal to $\mathcal{M}_{\mathfrak{D}}$ defined by Eq.~\eqref{equ.V-rep-def} with $\mrm{EP}(\mathfrak{D}) = \lrg{i}$ up to a phase, because a Majorana fermion operator squares to one, and thus the Majorana fermion operators in $\mathcal{M}_{\mathfrak{D}}$ annihilate each other, except for those at the endpoints. Therefore, 
\[
\rho_F = \frac{1}{2^{N/2}}\sum_{\mrm{EP}(\mathfrak{D})}
\eta_{\mathfrak{D}} \anglelr{\mathcal{M}_{\mathfrak{D}}}\mathcal{M}_{\mathfrak{D}}.
\]
Because $\eta_\mathfrak{D} \anglelr{\mathcal{M}_{\mathfrak{D}}} = \mathfrak{C}_{\mathfrak{D}}$ in the pure Kitaev model, and an endpoint set $\mrm{EP}(\mathfrak{D})$ uniquely corresponds to an equivalence class $[\mathfrak{D}]$, the DM can finally be written in the form of Eq.~\eqref{equ.rho-F}. Combining these results, we ultimately obtain the DM of the ground state
\begin{equation}\label{equ.Majorana-density-matrix}
    \rho = 2^{N-1}\mathcal{P}_G\prod_{\anglelr{ij}}\lr{\frac{\mathds{1} + \hat{u}_{ij}}{2}} \rho^F \mathcal{P}_G.
\end{equation}

Here, we also demonstrate that the DM in the spin representation can also be derived from the Majorana fermion representation \eqref{equ.Majorana-density-matrix}, We apply the following transformation
\[
\dimer{i}{j} = \mathcal{P}_G \hat{u}_{ij}ic_jc_i \mathcal{P}_G,
\]
to transform the Majorana fermion operators back into spin operators. We have applied the fact that $ib^\mu_ic_i$ commutes with the $\mathbb{Z}_2$ gauge generator $D_i$, thus $[ib^\mu_i c_i, \mathcal{P}_G]=0$. In general, the above relation implies the direct correspondence between $\Sigma_\mathfrak{D}$ and $\mathcal{M}_\mathfrak{D}$
\begin{equation}\label{equ.U-V-relation} 
    \mathcal{P}_G \lr{\prod_{\anglelr{ij}\in\mathfrak{D}}\hat{u}_{ij}} \mathcal{M}_\mathfrak{D} \mathcal{P}_G 
    = 
    \Sigma_{\mathfrak{D}}\mathcal{P}_G.
\end{equation}
The Gutzwiller projection of the last term is actually redundant since $\Sigma_{\mathfrak{D}}$ is already physical. We add this term here to emphasize that we restrict ourselves to the physical sector. With this equation, the gauge part, composed of gauge-field operators, can then be represented in terms of $\Sigma_\mathfrak{D}$ and $\mathcal{M}_\mathfrak{D}$. Therefore, Eq.~\eqref{equ.Majorana-density-matrix} can be rewritten as
\begin{align*}
    \rho &= \frac{2^{N-1}}{2^{3N/2}} \sum_{\mathfrak{D}} 
    \mathcal{P}_G \lr{\prod_{\anglelr{ij}\in\mathfrak{D}} \hat{u}_{ij}} 
    \rho^F 
    \mathcal{P}_G \\
    &=\frac{1}{2^{N+1}} \sum_{\mathfrak{D}} 
    \sum_{\mrm{EP}(\mathfrak{D}')} \mathfrak{C}_{\mathfrak{D}'}
    \mathcal{P}_G \lr{\prod_{\anglelr{ij}\in\mathfrak{D}} \hat{u}_{ij}} \mathcal{M}_{\mathfrak{D}'} \mathcal{P}_G
    .
\end{align*}
The denominator $2^{3N/2}$ in the first line arises from the gauge field projector and the fact that the number of edges of a honeycomb lattice is two-thirds the number of lattice sites. In the above derivation, the fermionic density matrix in the first line was substituted by Eq.~\eqref{equ.rho-F}.
The term sandwiched by the Gutzwiller projection is non-vanishing only if $[\mathfrak{D}] = [\mathfrak{D}']$. If the condition is satisfied, the term along with the Gutzwiller projection reduces to the string operator $\Sigma_{\mathfrak{D}}$ according to Eq.~\eqref{equ.U-V-relation}. As a result, we obtain the following DM
\begin{equation}\label{equ.spin-rho-from-fermion}
    \rho = \frac{1}{2^{N+1}} \sum_{\mathfrak{D}} \mathfrak{C}_\mathfrak{D} \Sigma_\mathfrak{D}. 
\end{equation}
The Gutzwiller projector has been omitted because we now completely restrict ourselves to physical space. The normalization factor is $1/2^{N+1}$ instead of $1/2^{N}$ because two different edge configurations $\mathfrak{D}_Z\ominus\mathfrak{D}$ and $\mathfrak{D}_{zz}\ominus\mathfrak{D}$ give the identical string operator $\Sigma_{\mathfrak{D}_Z\ominus\mathfrak{D}} = \Sigma_{\mathfrak{D}_{zz}\ominus\mathfrak{D}}$ for any $\mathfrak{D}$, as explained in Appendix~\ref{sec:fermion_parity_from_string_operators}. Thus, the result should be divided by two to avoid redundancy. Notice that we do not have this issue in $\rho^F$ since $\mathcal{M}_{\mathfrak{D}_Z}$ is not identical to $\mathcal{M}_{\mathfrak{D}_{zz}}$, which is proportional to identity $\mathcal{M}_{\mathfrak{D}_{zz}}\propto\mathds{1}$. Therefore, we have to treat them as two different components, and both should be included in $\rho^F$. After grouping the terms in Eq.~\eqref{equ.spin-rho-from-fermion} into equivalence classes, it will eventually reproduce Eq.~\eqref{equ.density-matrix}. 

\section{Proof of Eq.~\eqref{equ.Pick_sim-con}}\label{appendix:Pick}

\begin{figure}
    \centering
    \includegraphics[width=0.6\linewidth]{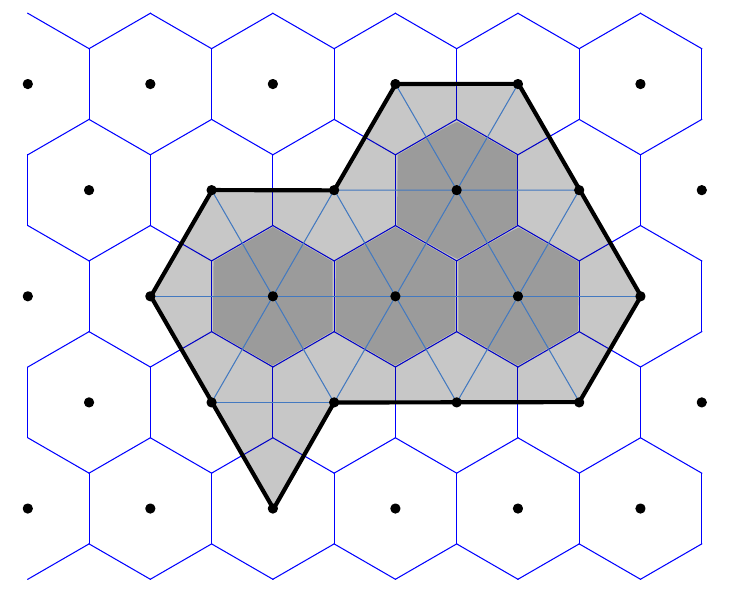}
    \caption{A triangular lattice and a honeycomb lattice are dual to each other, in the sense that each lattice point of a system lies at the center of a face of the other one. The boundary $C=\de A$ of the shaded region $A$ is a \textit{circuit} defined in the triangle lattice. $A$ consists of $f$ faces of the triangular lattice, where each face corresponds to a vertex of the dual honeycomb lattice, implying $f=N^A$. The darker region consists of $N^A_p$ hexagons entirely inside $A$, which equals the number of vertices $v$ of the triangular lattice enclosed by $C$. The length $\ell\equiv\abs{C}$ of the circuit can be understood as the number of edges in the honeycomb lattice that cross $C$.}
    \label{fig.dual-lattice-Pick-theorem}
\end{figure}

We find it more convenient to prove the formula by mapping the original honeycomb lattice to its dual triangular lattice, because a theorem on this problem has been discussed for triangular lattices~\cite{Lieb-Loss_1992}.

We begin by defining a \textit{circuit} $C$ of length $\ell$ in a triangular lattice as an ordered sequence of distinct sites $i_1,\cdots, i_\ell$ with the periodic property that $(i_n, i_{n+1})$ is an edge for $n = 1, \cdots, \ell$ with $i_{\ell+1} \equiv i_1$. The area enclosed by $C$ is denoted by $A$, which is simply-connected. Lemma 2.3 of~\cite{Lieb-Loss_1992} gives the following relation
\begin{equation}\label{equ.Lieb-Loss-formula}
    \ell = f - 2v + 2
\end{equation}
where $f$ and $v$ denote the number of interior \textit{faces} and \textit{vertices} of $A$, respectively. A proof of Eq.~\eqref{equ.Lieb-Loss-formula} can be derived from Euler's theorem, which states $2 = V-E+F$ for a convex polyhedron. Since a two-dimensional sublattice can be viewed as one side of a convex polyhedron, Euler's formula reduces to $2 = (\ell + v) - (\ell + e) + (f+1)$, where $e$ is the number of interior edges. Due to the geometry of a triangular lattice, by counting the faces and the edges they share, we obtain the relation $2e+\ell = 3f$. Substituting this relation into Euler's theorem, we obtain Eq.~\eqref{equ.Lieb-Loss-formula}. An illustration is given by \autoref{fig.dual-lattice-Pick-theorem}, which also shows that the dual lattice of a triangular lattice is a honeycomb lattice. This duality establishes correspondences between the triangular and honeycomb lattices: they are $\ell\leftrightarrow\abs{\de A}$, $f\leftrightarrow N^A$, and $v\leftrightarrow N^A_p$. According to these correspondences, replacing the quantities in Eq.~\eqref{equ.Lieb-Loss-formula} defined in a triangular lattice with their corresponding quantities defined in the dual honeycomb lattice gives Eq.~\eqref{equ.Pick_sim-con}.

\bibliography{ref}%
\bibliographystyle{my_apsrev4-2}

\end{document}